\documentclass[preprint,12pt]{elsarticle}
\usepackage[top=1in,bottom=1in,left=1in,right=1in]{geometry}

\usepackage{amssymb}
\newcommand{\Tr}{\mbox{Tr}}
\usepackage{amsthm, amsmath}

\usepackage{txfonts}
\usepackage{setspace}
\usepackage{xcolor}
\usepackage{caption}
\usepackage{subcaption}
\usepackage{float}
\usepackage[pdfborder={0 0 0},colorlinks,allcolors=blue]{hyperref}
\theoremstyle{definition}
\newtheorem{remark}{Remark}
\usepackage{comment}
\usepackage{xcolor, soul}

\definecolor{fgwhite}{rgb}{1,1,1}     
\definecolor{fgred}{rgb}{0.8,0,0}     
\definecolor{fgorange}{rgb}{0.93,0.53,0.18}     
\definecolor{fgpurple}{rgb}{0.55,0.1,0.6}     
\definecolor{fggreen}{rgb}{0,0.5,0}     
\definecolor{bggreen}{rgb}{0.8,1,0.8}     
\definecolor{fgblue}{rgb}{0,0,0.7}     
\definecolor{bgblue}{rgb}{0.9,0.9,1}     
\definecolor{fgclay}{rgb}{0.51,0.25,0.04}     
\definecolor{bggreen}{rgb}{0.8,1,0.8}     

\allowdisplaybreaks

\journal{-}

\begin{document}

\begin{frontmatter}

\title{Quantum-informed simulations for mechanics of materials: DFTB+MBD framework}

\author[inst1]{Zhaoxiang Shen}

\affiliation[inst1]{organization={Department of Engineering; Faculty of Science, Technology and Medicine; University of Luxembourg},
            city={Esch-sur-Alzette},
            postcode={4365}, 
            country={Luxembourg}}

\author[inst2]{Ra\'ul I. Sosa}
\author[inst1]{St\'ephane P.A. Bordas}
\author[inst2]{Alexandre Tkatchenko}
\author[inst1,inst3]{Jakub Lengiewicz}

\affiliation[inst2]{organization={Department of Physics and Materials Science, University of Luxembourg},
            city={Luxembourg City},
            postcode={1511}, 
            country={Luxembourg}}

\affiliation[inst3]{organization={Institute of Fundamental Technological Research, Polish Academy of Sciences},
            city={Warsaw},
            country={Poland}}

\begin{abstract}
The macroscopic behaviors of materials are determined by interactions that occur at multiple lengths and time scales. Depending on the application, describing, predicting, and understanding these behaviors require models that rely on insights from electronic and atomic scales. In such cases, classical simplified approximations at those scales are insufficient, and quantum-based modeling is required. In this paper, we study how quantum effects can modify the mechanical properties of systems relevant to materials engineering. We base our study on a high-fidelity modeling framework that combines two computationally efficient models rooted in quantum first principles: Density Functional Tight Binding (DFTB) and many-body dispersion (MBD). The MBD model is applied to accurately describe non-covalent van der Waals interactions. Through various benchmark applications, we demonstrate the capabilities of this framework and the limitations of simplified modeling. We provide an open-source repository containing all codes, datasets, and examples presented in this work. This repository serves as a practical toolkit that we hope will support the development of future research in effective large-scale and multiscale modeling with quantum-mechanical fidelity.
\end{abstract}


\end{frontmatter}


\section{Introduction}
\label{sec:Introduction}

The properties and behaviors of materials at the continuum scale are deeply rooted in the interactions and phenomena that take place at much smaller scales, both spatially and temporally. For example, the macro-scale plasticity and failure mechanisms in metals are significantly influenced by their microstructural properties and the dynamics of defects such as dislocations \cite{HU20123480,HOCHRAINER2014167}. Similarly, the macroscopic behavior of complex fluids is heavily dependent on their molecular structure \cite{weinan2011principles}, and the phenomena of adhesion and cohesion at interfaces are fundamentally linked to interactions at the micro and nanoscale \cite{vakis_modeling_2018}.
Moreover, for many important systems and applications, it is the quantum-scale effects that begin to show a noticeable influence at the macroscopic level. Such effects are observed when a multitude of atomic systems correlate to produce non-local collective phenomena, e.g., in superconductivity \cite{clarke2008superconducting}, photosynthesis \cite{photosynthesis}, in the principles behind quantum computing \cite{ladd2010quantum}, among others.

In scenarios where quantum effects play a pivotal role, precise macroscopic predictions demand the use of high-fidelity \emph{ab initio} models that are grounded in the fundamental principles of quantum mechanics. Among the \emph{ab initio} methods, Density Functional Theory (DFT) \cite{PhysRev.136.B864,PhysRev.140.A1133,RevModPhys.87.897} has gained prominence for its ability to offer high-fidelity results while maintaining a lower computational cost compared to other first principle methods, especially for systems comprising up to a few hundred atoms. DFT has found widespread application across various fields, including physics, chemistry, and material science, and has become a benchmark for the development of simplified or surrogate models that can tackle the complexity of material systems \cite{maurer2019advances, PhysRevLett.93.175503, curtin2003atomistic, REBO, SurfaceReacDFT, nyshadham2019machine, HOLLER2020103342}.

The computational cost of first principles/\emph{ab initio} methods, such as DFT, precludes their direct application to engineering-scale systems. In this context, they can be contrasted with \emph{fully empirical} methods, which rely on observations, parametrizations and experimental data to formulate models, rather than deriving interactions directly from fundamental theories. Fully empirical methods, such as molecular mechanics (MM) or force field (FF) methods, can thus be used to simulate systems millions of times larger. These approaches model atoms as classical particles, with the system described by a function of particle positions. The parameters and functional forms used to describe interactions between atoms and molecules are often derived from experimental measurements or quantum-mechanical calculations for small fragments. Despite the fact that empirical methods are widely employed for engineering purposes, they suffer from lower accuracy as compared to \emph{ab initio} methods. In fact, there is no systematically improvable way to ensure the reliability and predictive power of empirical force fields for complex molecules and materials. Hence, developing force fields with increasingly higher fidelity (towards the exact solution of the Schr{\"o}dinger equation) is the only guaranteed way to reach high-fidelity simulations of engineering-scale systems.

Bridging the gap between \emph{fully empirical} and \emph{ab initio} approaches, semi-empirical (SE) methods, such as the tight-binding method \cite{TightBinding_1997} and Hartree-Fock-based methods \cite{fischer1977hartree}, are able to balance efficiency and accuracy. SE methods incorporate empirical approximations into \emph{ab initio} methods to enhance their performance, enabling the study of larger systems with a controlled loss of accuracy. Among these methods, Density Functional Tight Binding (DFTB) \cite{Porezag199512947,Elstner19987260}, which is directly parameterized from DFT, emerges as a sound option due to its strong foundation in first principles. As a more computationally efficient tool, DFTB has been actively applied to large molecules (DNA, proteins, etc.), clusters, and nanoparticles (silver, gold, molybdenum disulfide, etc.), see \cite{DFTB_application_review} for a recent literature review of its applications. As such, it will be used in the present work as one of building blocks of the quantum-informed framework.

A usual limitation of DFT and DFT-based models, such as DFTB, is the absence of long-range correlation effects. This includes van der Waals dispersion forces, also known as London Dispersion Forces, in which electrons in one atom or molecule can influence the distribution of electrons in another, leading to correlated behavior.\footnote{In the context of quantum mechanics and computational chemistry, the term ``correlation'' refers to the mutual influence or connection between the motions of electrons in a system. Specifically, ``long-range correlation'' and ``long-range interactions'' implies interactions between electrons that extend over relatively large distances.}  This influence is not adequately captured by simpler models, and methods like DFT and DFTB may struggle to account for these correlations accurately. 
This can pose a significant drawback, as vdW dispersion interactions play a crucial role in numerous and widely different phenomena, from cohesive interactions in layered materials \cite{liu2019van} and protein folding \cite{ProteinFolding}, to the principles behind adhesion in gecko feet \cite{gecko}. Therefore, proper treatment of dispersion vdW interactions is needed to properly understand and predict properties of a multitude of engineering systems, which will also be demonstrated further in the present work.

Conveniently, vdW dispersion interactions can be added on top of existing DFT-based models, similarly as it is often done in the case of empirical methods.
Among the main approaches developed to incorporate vdW dispersion interactions, classical pairwise (PW) approximations, such as the Lennard-Jones (LJ) or Tkatchenko-Scheffler (TS) models \cite{PhysRevLett102073005}, have proven to be reliable for specific molecular systems and useful due to their computational efficiency. However, they neglect the quantum many-body nature of vdW interactions, which may prevent them from properly reproducing experimental observations \cite{Supramolecular,OrganicMolecularMaterials,Polymorphism}. For that reason, a more complex but also more accurate many-body dispersion (MBD) method \cite{PhysRevLett.108.236402,PhysRevLett102073005} for vdW interactions becomes preferable, as it takes into account collective long-range electron correlations. Numerous studies have convincingly shown that MBD surpasses PW models in accurately describing vdW dispersion interactions and more closely approaches experimental references, as seen in thin-layer delamination \cite{Hauseux}, vdW power law at the nanoscale \cite{Ambrosetti1171}, and modeling organic molecular materials \cite{OrganicMolecularMaterials}, among others.

Despite being more efficient than DFT and more accurate than empirical MM approaches, with notable examples being reported in the physics community \cite{DFTB+MBD_Mortazavi, Water_Martin}, the DFTB+MBD framework has not yet been widely used in the mechanical/computational engineering community. This is because the framework is relatively new to the engineering community, which is additionally exacerbated by the scarcity of accessible theoretical and practical introductions. It may be also unclear how certain engineering-scale systems are sensitive to quantum-scale effects predicted by the DFTB+MBD framework. Therefore, there is an urgent need to investigate and demonstrate performance of the framework on engineering systems, while developing friendly approaches to its application.

In the present paper, we address the aforementioned needs by introducing the following contributions:
\begin{itemize}
    \item An engineer-friendly introduction to the proposed quantum-based framework and related physics concepts, supported by the open-source library \cite{QuaCrepo1} containing all the models, examples, and datasets featured in this paper. This effectively transfers the seemingly unapproachable physics toolkit to mechanical/computational engineers.
    \item Identification of engineering systems whose mechanical properties require incorporating quantum effects. This serves as a practical guide, aiding the selection of suitable models for specific systems
    \item Demonstrating the dominance of DFTB+MBD over simplified modeling, by showcasing differences on various systems. 
\end{itemize}
In particular, we will introduce and explain the DFTB+MBD modeling framework, and study how it can be applied to various systems and problems, with an emphasis on how important high-fidelity modeling can be to accurately predict mechanical properties. We evaluate the performance of the framework for selected
molecular systems, including interacting carbon chains as the starting toy system, single-wall carbon nanotubes (SWCNT) as single molecules with
closed structures and robust mechanical properties, and Ultra High Molecular Weight Polyethylene (UHMWPE) as multi-extended-molecule systems held together by vdW forces.

The remainder of this paper is structured as follows: Section~\ref{sec:dft_based_modeling} introduces the current paradigm of atomistic modeling, namely DFT, followed by an introduction to DFTB as its high-fidelity approximations and how they lead to the necessity of implementing the energy range separation technique, which forms the basis of our framework. Section~\ref{sec:vdw_modeling} describes two vdW dispersion models to be implemented in conjunction with DFTB: one is based on the many-body quantum mechanical nature of the problem (MBD) and the other a pairwise approximation of the interaction (PW). Section~\ref{Sec:Applications} presents our open-source repository and demonstrates the framework's capabilities through benchmark applications on various systems. This includes static and dynamic studies on carbon chains, focusing on vdW effects, as well as mechanical tests on SWCNT and UHMWPE, where we explore the interplay between DFTB and vdW interactions in different nanostructures, emphasizing the significance of the quantum effects in predicting mechanical properties of materials relevant to engineering. Finally, Section~\ref{sec:conclusion} summarizes our contributions and discusses the potential of the proposed framework.

\section{Density Functional Theory-based modeling}
\label{sec:dft_based_modeling}

In this section, we will present the origin and reasoning behind the methodology of our frame-
work, which is built according to the energy range separation (ERS) approach. The concept of
ERS arises from the practical simplifications made when implementing DFT, which was developed to address the computational difficulties of directly solving the
Schrödinger equation. For that reason, we will first provide an informative introduction to DFT
in Section~\ref{sec: dft}, starting from the first-principles modeling, followed by presenting the semi-empirical
approximation of DFT, namely DFTB, which is practically employed in our proposed framework.
This introduction will assist in explaining the necessity of incorporating the concept of ERS and
thus add vdW dispersion corrections to DFT-based models, see Section~\ref{sec: E_SR E_LR} for a detailed
discussion.

\subsection{Density Functional Theory}
\label{sec: dft}
The behavior and properties of atomistic structures can be described by the many-body Schrödinger equation \citep{cohen1977quantum}. In the simple stationary (time-independent) scenario of $N$ particles labeled by $\boldsymbol r_i$, the Schrödinger equation takes the following form
\begin{equation}
\left[ \sum_{i=1}^N \left( - \frac{1}{2} \nabla^2_i \right) + \sum_{i=1}^N V^{\text{ext}} (\boldsymbol r_i) + \sum_{i<j}^N U(\boldsymbol{r}_i,\boldsymbol{r}_j) \right] \Psi =E \Psi.
    \label{eq:Sch}
\end{equation}
In the equation above, the term in the square brackets is the Hamiltonian operator that includes the non-interacting term $V^{\text{ext}}(\boldsymbol r_i)$ (e.g., the external potential) and the interacting term $U(\boldsymbol{r}_i,\boldsymbol{r}_j)$, while $E$ is the unknown energy, and $\Psi=\Psi(\boldsymbol{r}_1,\boldsymbol{r}_2,\ldots,\boldsymbol{r}_N)$ is the unknown $3N$-dimensional complex-valued wave function that fully defines the state of the whole system.

Solving the many-body Schrödinger equation requires finding the eigenvalues, $E_k$, and eigenvectors, $\Psi_k$, of the Hamiltonian operator. This can be a challenging task for large, complex systems, since the memory requirement to store the whole system wave-function, as well as the computational cost it takes to calculate it, scales exponentially with the number of interacting particles.
For that reason, methods for approximating the quantum mechanical interactions are usually used. 

The most widely used simplifying step is the Born-Oppenheimer approximation for atom nuclei \citep{Szabo1996},
which is based on the fact that the nuclear polarization is much smaller than the electronic polarization.
This enables us to regard the atomic nuclei as classical particles, focusing solely on the quantum electronic problem—the interaction among $N$ electrons in the presence of nuclei. However, this approach proves insufficient for most practical applications, necessitating additional approximations due to the complexity of the full many-body problem involving all remaining electronic degrees of freedom, which is challenging to solve entirely. 

A commonly used method to mitigate the aforementioned complexity issue is the DFT method, which allows for coarse-graining from a 3N-dimensional wave function of interacting electrons, $\Psi(\boldsymbol{r}_1,\ldots,\boldsymbol{r}_N)$, to a much simpler 3-dimensional density function, $n(\boldsymbol{r})$. The DFT method is based on the Hohenberg-Kohn theorems \citep{hohenberg1964inhomogeneous},
which claim that the ground state solution for the general problem can be exactly represented as a system of $N$ non-interacting electrons subjected to an effective potential determined by the total ground state electron density, $n(\boldsymbol{r})$. 

The key point of considering electrons as non-interacting lies in the fact that it greatly simplifies the computational problem. By treating electrons as non-interacting, we can avoid the need to account for the complex many-body interactions between them, which would otherwise require a significantly higher computational effort. This qualitative change in the expression leads to an extreme reduction in computational cost. In fact, the cost of solving Eq.~\eqref{eq:Sch} for an unknown $3N$-dimensional wavefunction scales as 
$O(K^{3N})$ for $K$ equal to the number of discretization points used to compute it, while the cost of performing DFT scheme for an unknown 3-dimensional density function scales as 
$O(N^3)$. 

In DFT, the total energy $E$ can be written as a function of the electron density function $n(\boldsymbol{r})$ and atomic (nuclei) position $\boldsymbol{R}$,
\begin{equation}
\begin{split}
E\!\left(n(\boldsymbol{r}),\boldsymbol{R}\right) &= T\!\left(n(\boldsymbol{r})\right) + E^\text{ext}\!\left(n(\boldsymbol{r}),\boldsymbol{R}\right) + \frac{1}{2} \int \frac{n(\boldsymbol{r})n(\boldsymbol{r}')}{|\boldsymbol{r}-\boldsymbol{r}'|} d\boldsymbol{r}d\boldsymbol{r}' \\
&+ E^\text{xc}\!\left(n(\boldsymbol{r})\right) + \frac{1}{2} \sum_{i,j} \frac{Z_i Z_j}{|\boldsymbol{R}_i- \boldsymbol{R}_j|} ,
    \label{eq:E-Tot-DFT}
\end{split}
\end{equation} 
and is composed of five characteristic terms.
The first one is the total electronic kinetic energy $T\!\left(n(\boldsymbol{r})\right)$, the second represents the energy contribution due to the external potential induced by atomic charges $E^\text{ext}\!\left(n(\boldsymbol{r})\right)$, the third is called the Hartree term and it describes the Coulomb repulsion between electrons, the fourth is the exchange-correlation energy $E^\text{xc}\!\left(n(\boldsymbol{r})\right)$, and the last one is the atomic energy (nuclei-nuclei interaction). The exchange-correlation part is arguably the most important since it encompasses all of the characteristic quantum mechanical effects, including the many-body particle interactions.

The electron density function, $n(\boldsymbol{r})$, used in Eq.~\eqref{eq:E-Tot-DFT}, is obtained as a solution to a separate problem, the Konh-Sham equations \cite{KS-PhysRev.140.A1133}. The Konh-Sham equations are derived through the DFT formalism, which provides a non-interactive system description for electrons using the Kohn-Sham orbitals $\phi_k (\boldsymbol{r})$:
\begin{equation}
\left[ - \frac{1}{2} \nabla^2 + V^\text{ext} + \int \frac{n(\boldsymbol{r}')}{|\boldsymbol{r}-\boldsymbol{r}'|} d\boldsymbol{r}'+ \frac{\delta E^\text{xc}}{\delta n(\boldsymbol{r})}\right] \phi_k (\boldsymbol{r}) = \epsilon_k \phi_k (\boldsymbol{r}).
    \label{eq:KS-DFT}
\end{equation}
The four terms in this equation are directly related to the first four terms of Eq.~\eqref{eq:E-Tot-DFT}, and pose an eigenvalue problem to be solved for the unknown Kohn-Sham orbitals $\phi_k (\boldsymbol{r})$ and the related energies $\epsilon_k$.

In Eq.~\eqref{eq:KS-DFT}, the Kohn-Sham orbitals are represented by a linear combination of respective basis functions $\varphi_{l}(\boldsymbol{r})$, and their choice is specific to the problem in hand, see \cite{Parr1994}. As a result, $\phi_k(\boldsymbol{r})=\sum_{l}C_{rl}\varphi_{l}(\boldsymbol{r})$, with $C_{rl}$ being effectively the unknowns in Eq.~\eqref{eq:KS-DFT}. Knowing the orbitals, $\phi_k(\boldsymbol{r})$, it is possible to compute the electron density 
\begin{equation}
n(\boldsymbol{r}) = \sum_k |\phi_k(\boldsymbol{r})|^2,
    \label{eq:density-DFT}
\end{equation}
and solve the Eq.~\eqref{eq:KS-DFT} and \eqref{eq:density-DFT} in an iterative self-consistent manner. The only constraint is that the integral of the electron density $n(\boldsymbol{r})$, which represents the total charge, be preserved. 

The algorithmic process for solving the Kohn-Sham equations involves the following steps:
\begin{enumerate}
    \item Initialize the electron density $n(\boldsymbol{r})$ with an initial guess.
    \item Compute the last three terms in the bracket in Eq.~\eqref{eq:KS-DFT} using the current electron density $n(\boldsymbol{r})$.
    \item Solve the Kohn-Sham equations (Eq.~\eqref{eq:KS-DFT}) in order to find the eigenfunctions (orbitals) $\phi_k (\boldsymbol{r})$.
    \item Update the electron density $n(\boldsymbol{r})$ using the new orbitals $\phi_k (\boldsymbol{r})$ according to Eq.~\eqref{eq:density-DFT}.
    \item Check for convergence of the electron density $n(\boldsymbol{r})$. If not converged, return to step 2.
\end{enumerate}
Upon properly obtaining the electron density, we can derive the total energy precisely using Eq.~\eqref{eq:E-Tot-DFT}. However, even for systems with thousands of atoms, the computational cost of DFT makes it prohibitive for solving the energy minimization problem in a successive manner for exploring different configurations. In the following section, we will present an efficient high-fidelity alternative for atomistic simulations, known as DFTB. 

\begin{remark} It is important to note that the electron density $n(\boldsymbol{r})$ has an implicit dependency on the atomic positions $\boldsymbol{R}_i$, as the external potential $V^\text{ext}$ and the exchange-correlation functional $\frac{\delta E^\text{xc}}{\delta n(\boldsymbol{r})}$ depend on the positions of the nuclei. This dependency is properly resolved when computing the forces as a derivative of energy with respect to the atomic positions \cite{payne1992}. However, the discussion of particular approaches for doing so is beyond the scope of the present paper.
\end{remark}

\subsection{Density Functional Tight Binding} \label{Sect:DFTB}
Density Functional Tight Binding (DFTB) \citep{dftb2} is a semi-empirical method rooted in DFT that relies on the, so-called, tight binding approximation. The fundamental assumptions of this approach include the confinement of valence electrons
to their respective atoms and the representation of density fluctuations as a superposition of atomic contributions, which are considered to be exponentially decaying, spherically symmetric charge densities. In this study, we utilize the DFTB3 method \citep{dftb3}, a variant of DFTB that incorporates a third-order Taylor expansion of these tightly bound electronic states with respect to electron density fluctuations in the total energy calculation shown in Eq.~\eqref{eq:E-Tot-DFT}, based on the Kohn-Sham formalism introduced in Section~\ref{sec: dft}. Consequently, the total electron density $n(\boldsymbol{r})$ can be approximated using a linear combination of a minimal atomic basis set obtained through parametrization. In essence, the electron density of the system $n(\boldsymbol{r})$ is represented by combining the contributions of a limited set of atomic orbitals, which are fitted to reference data acquired from DFT calculations. This process allows for rewriting the Kohn-Sham equations (Eq.~\eqref{eq:KS-DFT}) in terms of the defining coefficients for this minimal atomic basis set, and solve a reduced eigenvalue problem to compute $n(\boldsymbol{r})$ following the same iterative procedure detailed in Section~\ref{sec: dft}.
Although the computational complexity of DFTB3 scales the same as DFT ($O(N^3)$) because it also relies on the diagonalization operation, the approximations employed in this case enable a speedup of one to two orders of magnitude compared to the latter while preserving accuracy for a broad range of systems with localized electrons. 

While DFTB3 offers a significant computational advantage over DFT, we should acknowledge its limitations as well as the range of systems for which it is well-suited. DFTB3 relies on an empirical parametrization, which can limit its accuracy and transferability compared to DFT. Additionally, its minimal basis set further reduces its reliability compared to DFT. Despite these limitations, DFTB3 has demonstrated success in accurately modeling various systems, such as organic molecules, molecular crystals, surfaces, and nanomaterials. For cases where DFTB is not well suited, such as systems characterized by presenting charge transfer or excited states,  alternative methods, including DFT itself, wavefunction-based approaches, or hybrid methods, provide much more accurate results.

\begin{remark} DFTB3 enables calculations on periodic lattices, which is utilised in some of the presented examples. The handling of periodicity in DFTB3 is facilitated by Bloch's theorem, which suggests that solutions to the Schrödinger equation for a periodic potential can be depicted as a product of a plane wave and a function that mirrors the periodicity of the lattice \cite{ashcroft1976}. This theorem allows the problem to be transformed into the reciprocal space, or k-space. In this context, k-points are specific points in the reciprocal space that represent different possible states for an electron in the crystal. For each of these k-points, the Kohn-Sham equations, a set of eigenvalue equations, need to be solved. This involves a process of diagonalization, resulting in a set of eigenvalues (the energy levels) and eigenvectors (the corresponding wave functions) for each k-point. The choice of k-points for the calculations can significantly impact both the accuracy and computational cost of the simulation. A larger number of k-points generally leads to more precise results, but also increases the computational cost due to the diagonalization process required for each k-point.
\end{remark}

\subsection{Energy range separation}
\label{sec: E_SR E_LR}
In DFT, the exact functional for the exchange-correlation is not known, which forces us to make simplifications to $E^\text{xc}$. The nature of these approximations is widely taken as local, i.e., we assume the exchange-correlation to depend only on the density at the coordinate where it is being evaluated or its low-order derivatives.  This assumption inevitably makes this interaction energy decay exponentially with distance, since that is the decay rate for the electron wavefunction overlap. Moreover, a second-order expansion analysis on the correlation energy \cite{doi:10.1063/1.4789814} reveals that, for two neutrally charged spherical particles, the energy decay should follow a rate inversely proportional to the distance to the power of $6$ generated because of induced instantaneous dipole interactions. This tells us that the interaction decay rate for DFT is overestimated and thus it is necessary to correct for the long-range correlation interactions by introducing a van der Waals dispersion potential. 

\begin{figure}[h]
\centering
\includegraphics[trim = 50mm 25mm 50mm 25mm, clip=true,width=0.6\textwidth]{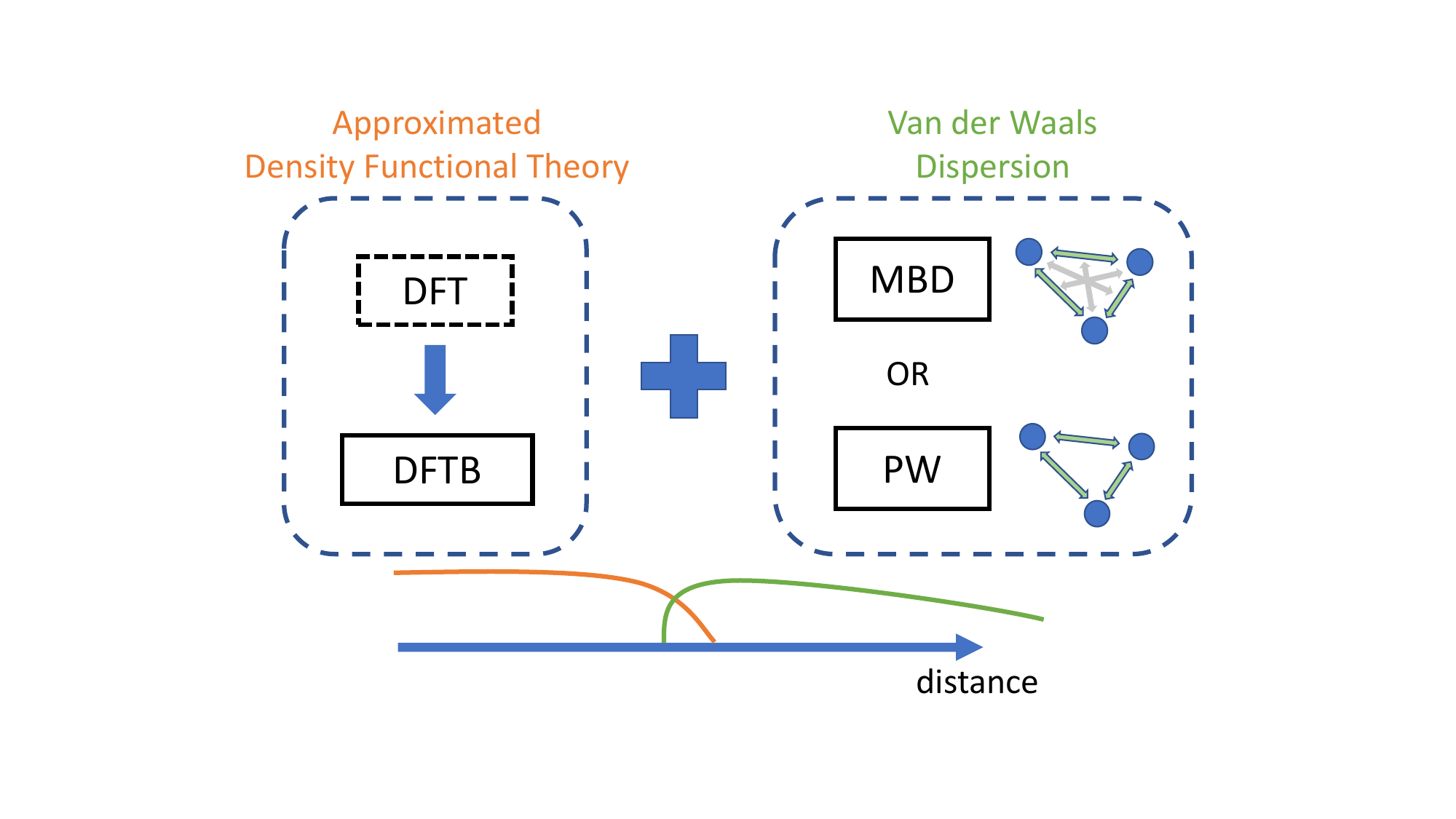}
    \caption{Illustration of the proposed high-fidelity framework and its energy range separation approach. The modeling is based on the approximated density functional theory (ADFT), for which we apply DFTB as an efficient semi-empirical approximation of DFT. The long-range complement for the ADFT is achieved by van der Waals dispersion interactions, and our framework offers both the many-body (MBD) and pairwise (PW) methods. The concept of energy range separation is abstracted as the interplay between the orange (ADFT) and green (vdW) curves, with a clear overlapping revealing the implicit nature of the separation.}
     \label{fig:framework_diagram}
\end{figure}

It is in this context that we introduce the energy range separation, see Fig.~\ref{fig:framework_diagram}, to decouple the total energy of an atomic system $E^\text{tot}$
\begin{equation}
    E^\text{tot} = E^\text{ADFT} + E^\text{vdW},  \label{eq:E_SR_LR_DF_MF}
\end{equation}
where the total energy is defined as the sum of the contribution obtained as a part of the approximated DFT energy $E^\text{ADFT}$, and the long-range contribution determined by van der Waals dispersion interactions $E^\text{vdW}$ which will be introduced in Section~\ref{sec:vdw_modeling}. 
In general, there are various options available for $E^\text{ADFT}$, each with different trade-offs between efficiency and accuracy. In our DFTB+MBD framework, we incorporate DFTB method to achieve a good balance between high-fidelity and computational feasibility for the non-vdW contribution in Eq.~\eqref{eq:E_SR_LR_DF_MF}. This can be expressed as 
\begin{equation}
    E^\text{tot} = E^\text{DFTB} + E^\text{vdW},
    \label{eq:E_SR_LR_DFTB}
\end{equation}
where vdW superscript represents either MBD or PW models.

\section{Van der Waals dispersion: pairwise and many-body models} 
\label{sec:vdw_modeling}
Van der Waals dispersion, also know as London dispersion, is a type of ubiquitous attractive inter-atomic/molecular force. It arises due to the temporary fluctuations in the electron density, leading to instantaneous dipoles which then interact with other dipoles induced by them. As stated in Section~\ref{sec:dft_based_modeling}, vdW dispersion interactions are not taken into account in the usual approximations of exchange-correlation energy made in DFT and DFTB methods, but can be included as an additional term, $E^{\text{vdW}}$, in the total energy. In this section, we will present two such vdW models, employed in our framework, i.e. pairwise and many-body dispersion approaches.

\subsection{Pairwise approach}
\label{sec:PW}
In order to model vdW interactions, pairwise approaches are commonly used. In those cases the vdW dispersion energy $E^{\mathrm{vdW, PW}}$ is evaluated as a simple sum of interactions between two atoms:
\begin{equation}
E^{\mathrm{vdW, PW}} = -\sum_{j>i}f^\text{damp} \cdot \dfrac{C_{6,ij}}{R_{ij}^{6}},
\label{eq:pw}
\end{equation}
where $R_{ij}=\|\boldsymbol{R}_{i}-\boldsymbol{R}_{j}\|$ is the interatomic distance between atoms $i$ and $j$. $C_{6,ij}$ are coefficients, which are either determined experimentally or numerically from DFT calculations. The damping function $f^\text{damp}$ prevents a non-physical singularity when $R_{ij}$ tends to zero and is equal to one at large distances. The properties at medium and small distances will be always slightly perturbed by Eq.~\eqref{eq:pw}, depending on the parameters of the damping function.

In our simulations, we adopt the Tkatchenko-Scheffler (TS) vdW model \cite{PhysRevLett102073005}. It is based on a scaling technique that determines accurate \textit{in situ} polarizabilities and other corresponding parameters by taking into account the local chemical environment. The damping function in the TS model is of Fermi-type defined as follows:
\begin{equation}
    f^\text{damp}(R_{ij}) = \left[ 1 + \exp(-d\cdot (R_{ij}/S^\text{vdW}-1)) \right]^{-1},
\label{eq:fdamp}   
\end{equation}
where $d$ is the parameter for adjusting the shape of the damping function. $S^\text{vdW}$ is defined as a scaled sum of vdW radii $\gamma(R_{i}^\text{vdW,eff}+R_{j}^\text{vdW,eff})$, where $R_{i}^\text{vdW,eff}$ is the effective vdW radius of the ith atom and $\gamma$ is a functional-specific empirical parameter that is fitted to reproduce intermolecular interaction energies based on the choice of experimental or benchmark database \cite{ PhysRevLett102073005,ts_database, ambrosetti2014long}. The pairwise coefficient $C_{6,ij}$ is defined as:
\begin{equation}
    C_{6,ij} = \frac{2C_{6,ii}^\text{eff} C_{6,jj}^\text{eff}}{(\alpha_{j}^\text{0,eff}/\alpha_{i}^\text{0,eff})C_{6,ii}^\text{eff}+(\alpha_{i}^\text{0,eff}/\alpha_{j}^\text{0,eff})C_{6,jj}^\text{eff}}
\end{equation}
where $\alpha_{i}^\text{0,eff}$ is the effective atomic polarizability, and $C_{6,ii}^\text{eff}$ is the effective atomic parameter. The above ``effective'' terms can be obtained by scaling their free-atom values, i.e.
\begin{equation}
    C_{6,ii}^\text{eff} = C_{6,ii}^\text{free} \left(\dfrac{V_i^{\mathrm{eff}}}{V_i^{\mathrm{free}}}\right)^2, \quad
    \alpha_{i}^\text{0,eff} = \alpha_{i}^\text{0,free} \left(\dfrac{V_i^{\mathrm{eff}}}{V_i^{\mathrm{free}}}\right), \quad
   R_{i}^\text{vdW,eff} = R_{i}^\text{vdW,eff} \left(\dfrac{V_i^{\mathrm{eff}}}{V_i^{\mathrm{free}}}\right)^{1/3},
\end{equation}
where the factor $V_i^{\mathrm{eff}}/V_i^{\mathrm{free}}$ represents the ratio of the effective volume occupied by the atom interacting with its environment, as evaluated using the Hirshfeld partitioning \cite{Hirshfeld}, to the free (non-interacting) reference volume occupied by this atom. In general, the volume ratios vary throughout an atomic structure, which can only be retrieved from high fidelity models, such as the DFTB method introduced in Section~\ref{sec:dft_based_modeling}.

\subsection{Many-body dispersion approach}
\label{sec:MBD}
The MBD method \cite{PhysRevLett.108.236402,doi:10.1063/1.4789814} is a powerful approach for calculating the interatomic interaction energy based on the Adiabatic Connection Fluctuation Dissipation Theorem (ACFDT) within the Random Phase Approximation (RPA) for a model system comprised of quantum harmonic oscillators (QHO) interacting via the dipole-dipole interaction potential \cite{proof2013jcp}. The MBD model generalizes the pairwise vdW energy expression by considering all orders of the dipole interaction between fluctuating atoms. Following \cite{ambrosetti2014long}, the ACFDT-RPA correlation energy for MBD model is expressed as follows:
\begin{equation}
E^{\textrm{vdW,MBD}} = \dfrac{1}{2\pi} \int _0 ^\infty \Tr \left[ \mathrm{ln}(\boldsymbol{1} - \boldsymbol{AT})\right] \text{d} \omega,
\label{eq:MBDEnergy}
\end{equation}
where $\boldsymbol{A}$ is a diagonal $3N \times 3N$ matrix ($N$ atoms). For the case of isotropic QHOs, $\boldsymbol{A}$ is defined as ${A}_{lm} = -\delta_{lm} \alpha_l \left( i \omega \right)$, where  $\alpha_l \left( i \omega \right)$ is the $l$th frequency-dependent atomic polarizability. $\boldsymbol{T}$ is the dipole-dipole interaction tensor:
\begin{equation}
T_{ij}^{ab} = \nabla_{\boldsymbol{R}_i} \otimes \nabla_{\boldsymbol{R}_j} v_{ij}^{gg},
\label{ddinter}
\end{equation}
where $a$ and $b$ specify the Cartesian coordinates, and $v_{ij}^{gg}$ is a modified Coulomb potential \cite{KWIK_erf}, used to incorporate overlap effects for a set of fluctuating point dipoles:
\begin{equation}
v_{ij}^{gg} = \dfrac{\text{erf}\left(  R_{ij}/ (\beta\cdot \tilde{\sigma}_{ij})  \right) }{R_{ij}},
\label{ModiCoulPo}
\end{equation}
in which $\beta$ in Eq.~\eqref{ModiCoulPo} is an empirical constant, $R_{ij}$ is the interatomic distance between atom $i$ and $j$, and $\tilde{\sigma}_{ij} = \sqrt{\sigma_{i}^2 + \sigma_{j}^2}$ is an effective width computed from the atom's Gaussian widths $\sigma_{i}$ and $\sigma_{j}$ of atoms $i$ and $j$ respectively. These Gaussian widths are directly related to the polarizabilities, $\alpha_{i}$, in classical electrodynamics \cite{PhysRevB.75.045407} and can be derived from the dipole self-energy as:
\begin{equation}
\sigma_{i} = \left(\sqrt{\dfrac{2}{9\pi}} \alpha_{i}\right)^{1/3}. 
\label{eq:Gaussian_width}
\end{equation}

For the efficiency of implementations, we assume the dipole-dipole interaction is frequency-independent by utilizing the effective polarizability $\alpha_{i}^{\text{0,eff}}$ to compute $\sigma_{i}$ as defined by Eq.~\eqref{eq:Gaussian_width}. According to \cite{ambrosetti2014long}, the ACFD-RPA correlation energy defined by Eq.~\eqref{eq:MBDEnergy} can be directly obtained (for a frequency-independent $\boldsymbol{T}$ tensor) from diagonalization of a model Hamiltonian that is based on atom-centered QHOs coupled by dipole-dipole interactions. The interaction energy can thus be directly evaluated from the 3$N$ (for $N$ atoms) eigenvalues $\lambda_p$ of the matrix composed by $N^2$ $3 \times 3$ blocks $\boldsymbol{C}_{ij}^{\mathrm{MBD}}$ which characterize the coupling between each pair of atoms $i$ and $j$: 

\begin{equation}
    \boldsymbol{C}_{ij}^{\mathrm{MBD}} = \omega_i^2 \delta_{ij} + (1-\delta_{ij}) \ \omega_i\omega_j \sqrt{ \alpha_{i}^\text{0,eff} \alpha_{j}^\text{0,eff}}\ \boldsymbol{T}_{ij}.
    \label{eq:Matrix E_mbd}
\end{equation}
where $\omega_i$ is the characteristic frequency for atom $i$ that is determined as follows \cite{PhysRevLett102073005}:
\begin{equation}
\omega_i = \dfrac{4\ C_{6,i}^\text{free}}{3\ (\alpha_i^\text{0,free})^2}.
\end{equation}
The MBD energy is finally computed as the difference between the interacting and non-interacting frequencies:
\begin{equation}
E^{\text{vdW,MBD}} = \dfrac{1}{2} \sum_{p=1}^{3N} \sqrt{\lambda_p} - \dfrac{3}{2} \sum_{i=1}^{N} \omega_i. 
\label{eq:mbd_energy_QHO}
\end{equation}
The interatomic forces can be computed by directly taking the gradient of the energy defined in Eq.~\eqref{eq:mbd_energy_QHO} instead of the more complex and equivalent expression given in Eq.~\eqref{eq:MBDEnergy}. Hence, the 3-dimensional force vector of atom $i$ is obtained as: $\boldsymbol{F}_i^{\mathrm{MBD}}=-\nabla_i E^{\text{vdW,MBD}}$, where the gradient is taken w.r.t. $\boldsymbol{R}_i$. After some algebra one finds
\begin{equation}
\boldsymbol{F}_i^{\rm MBD}=-\frac{1}{4} \Tr \left[ \boldsymbol{\Lambda}^{-1/2}\, \boldsymbol{S}^\text{T} \left(\nabla_i\boldsymbol{C}^{\mathrm{MBD}}\right)\boldsymbol{S}\right],\\
\label{eq:F_mbd}
\end{equation}
where $\boldsymbol{\Lambda}_{pq}=\lambda_p\delta_{pq}$ is obtained upon diagonalization of $\boldsymbol{C}^{\mathrm{MBD}}$ via a suitable SO (3$N$) rotation matrix $\boldsymbol{S}$ that diagonalizes $\boldsymbol{C}^{\mathrm{MBD}}$: $\boldsymbol{\Lambda}=\boldsymbol{S}^\text{T}\boldsymbol{C}^{\mathrm{MBD}}\boldsymbol{S}$. 
The trace operator implies summation over the $3N$ eigenmodes of the system, each reflecting a collective contribution from the whole system. This reveals a qualitative difference from the PW models, in which the force split has a pairwise nature.

\begin{remark}
In the following applications which employ DFTB, a variant of the MBD method, namely MBD@rsSCS \cite{ambrosetti2014long}, is applied. This variant follows the structure introduced above, however, it is adapted to be used with DFT-based models such as DFTB. For instance, it is capable of capturing the so called screening effect via self-consistent screening, which affects dipolar fluctuations that induce vdW dispersion \cite{PhysRevLett.108.236402}.

\end{remark}


\section{Applications}
\label{Sec:Applications}

In this comprehensive section, we will demonstrate the modeling capabilities of the presented DFTB-based framework. In Section~\ref{Sect:Num-imp} we provide implementation notes and give an overview of the open-source repository being a part of this contribution, aiming to make the DFTB-based framework more easily approachable for the engineering community. Following that, through a series of benchmark examples of increasing complexities, we will try to shed light on how quantum effects can impact the mechanical properties of certain engineering systems. In Section~\ref{sec:vdW_chain}, benchmark static and dynamic calculations are performed on interacting carbon chains, to demonstrate quantum many-body vdW effects on simple and flexible structures. Subsequently, in Section~\ref{sec:SWCN}, we shift our focus to the single-wall carbon nanotube (SWCNT), a much stiffer benchmark system, and conduct buckling tests to emphasize the necessity of high-fidelity modeling in accurately capturing mechanical responses, while illustrating the pitfalls of simplified modeling. Finally, in Section~\ref{sec:UHMWPE}, inspired by the insights gained from the first two benchmark cases, we arrive at the application on a multi-extended-molecule system, namely Ultra High Molecular Weight Polyethylene (UHMWPE). The mechanical properties of these systems are known to rely on vdW interactions, and we show that the interplay between the DFTB and MBD components of the framework is necessary to model these effects accurately.

\subsection{Implementation notes} 
\label{Sect:Num-imp}

In the examples studied in the following sections, we utilize models provided by two software packages: the established DFTB+ library and the novel TensorFlow-based library. A concise introduction to both is provided subsequently. For the benefit of future research, we offer an open-access repository \cite{QuaCrepo1} that encompasses all the software, numerical examples, and datasets utilized in this study.
The first of the libraries, the DFTB+ software package \cite{DFTB+}, represents an established implementation of the DFTB+MBD framework, as introduced in Sections~\ref{sec:dft_based_modeling} and~\ref{sec:vdw_modeling}. This package utilizes the DFTB3 method and incorporates the vdW dispersion interactions exclusively through the Fortran-based library \emph{libmbd} \cite{libmbd-code}. Additionally, it offers essential tools for conducting molecular dynamics simulations and structural optimization. To enhance the accessibility of the DFTB+ library to the mechanical engineering community, we have provided a wrapper in the form of a Python application programming interface (API). This API facilitates seamless importing and manipulation of geometry, modification of simulation parameters, execution of intricate static and dynamic tests, and post-processing of output data, extending beyond the inherent capabilities of DFTB+. With this toolkit, users can undertake comprehensive analyses using a single script file, enhancing both convenience and efficiency.

The second library utilized in this work was developed as an alternative programming approach that leverages the TensorFlow (TF) Python library. The primary motivation behind this choice is TF's automatic differentiation (AD) capabilities, which facilitate the development of new models in a quasi-symbolic manner, thus obviating the need for low-level coding of gradients and Hessians. Another significant advantage of using TF is its potential for seamless integration with cutting-edge acceleration techniques, including machine-learning surrogate modeling and GPU acceleration, positioning it as a favorable avenue for future advancements. In this paper, we have implemented various vdW dispersion interaction models, such as the MBD and PW models discussed in Section~\ref{sec:vdw_modeling}. For modeling short-distance interactions (e.g., covalent bonds), we employ the TF framework to implement the simplified harmonic model, as showcased in Section~\ref{sec:harmonic}, enabling us to examine the constraints of such simplified modeling.

\subsection{vdW effects on simple structures}
\label{sec:vdW_chain}

\begin{figure}[h]
    \centering
    \includegraphics[trim = 25mm 18mm 0mm 215mm, clip=true,width=0.65\textwidth]{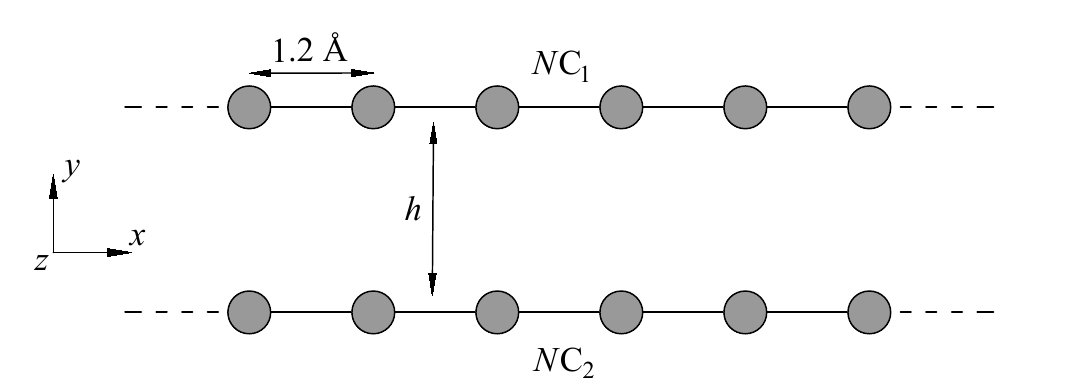}
    \caption{Schematic of the system which includes 2 parallel chains with an inter-chain distance $h$. The upper and lower chains contain $N\text{C}_1$ and $N\text{C}_2$ atoms, respectively, while the spacing between neighboring carbon atoms is $1.2\,\text{\AA}$.}
    \label{fig:carbon_chains}
\end{figure}

In this section, we analyze simple systems of two interacting carbon chains, as shown in Fig.~\ref{fig:carbon_chains}, and compare predictions between two vdW models introduced in Section~\ref{sec:vdw_modeling}, i.e., the MBD method and the pairwise TS model (PW in short).  We study three cases of increasing complexity: the fully rigid case, the flexible quasi-static case, and the fully dynamic case. Each of those settings allows us to demonstrate specific effects, and consequently to better understand the differences between the two vdW models.

\begin{figure}[t]
    \centering
    \subfloat[Total vdW force decay with distance.]{\includegraphics[trim = 28mm 90mm 30mm 90mm, clip=true,width=0.49\textwidth]{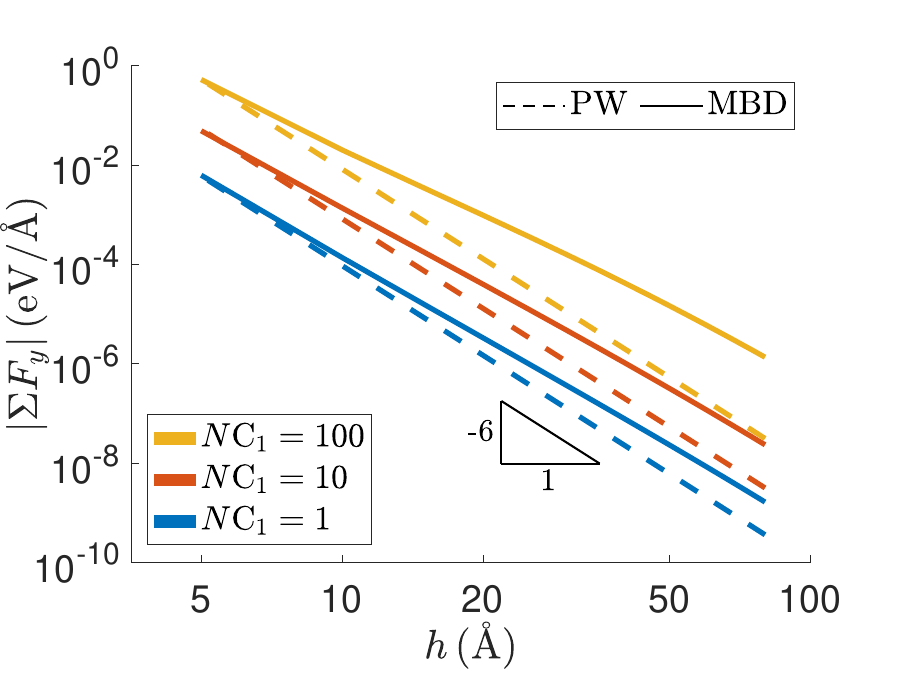}\label{fig:MBDvsPW_chains_h}}\hfil
    \subfloat[Total vdW force change by size.]{\includegraphics[trim = 30mm 90mm 30mm 90mm, clip=true,width=0.49\textwidth]{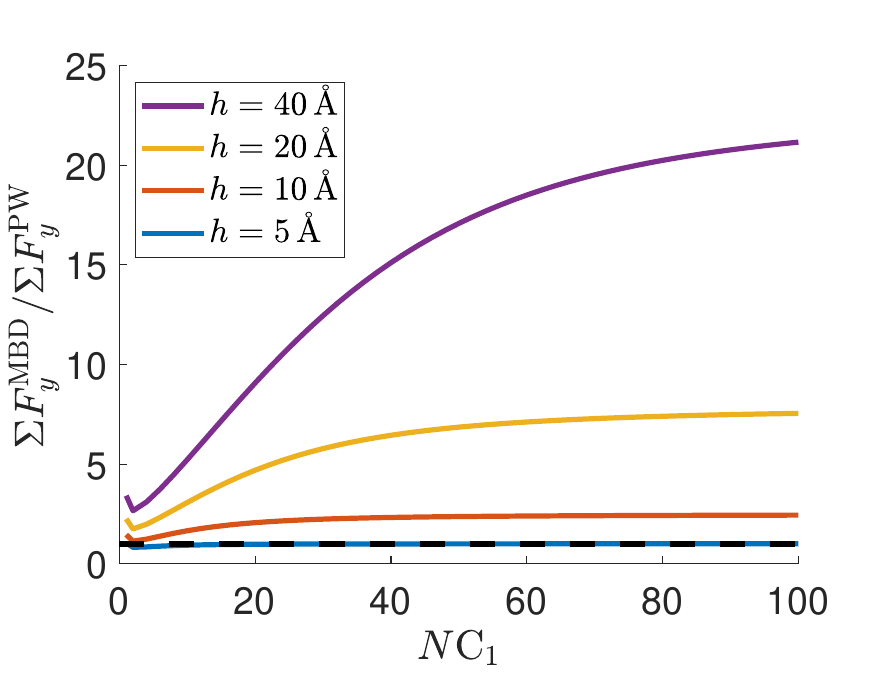}\label{fig:MBDvsPW_chains_N}}
    \caption{Static vdW force calculation on the carbon 2-chain system. (a) shows the scaling of the net vdW force in $y$ direction by the changing $h$. (b) shows the evolution of the net force ratio of MBD to PW when changing length of upper chain $N\text{C}_{1}$.
    }
    \label{fig:MBDvsPW_chains}
\end{figure}

In the rigid case, we study the vdW net forces between two fixed parallel carbon chains for varying distance, ${h}$, between the chains, and a varying number of atoms, $N\text{C}_1$, in the upper chain. The lower chain is kept relatively long, $N\text{C}_2=200$, approximating an infinite substrate. The results shown in Fig.~\ref{fig:MBDvsPW_chains} reveal significant relative differences in interaction forces, $\sum\!F_y^{\,\text{PW}}$ and $\sum\!F_y^{\,\text{MBD}}$, predicted by PW and MBD models, respectively. While at short distances the net forces predicted by both vdW models are comparable, the MBD force decays slower with distance, which is additionally amplified with the increasing number of atoms in the chains. This evident and strong many-body effect relies on the quantum nature of dispersion interactions, which can be accurately predicted by the MBD method, and can not be captured by pairwise models that are by definition additive and assume a constant decay rate (see Eq.~\eqref{eq:pw}).

The enhancement in the total interaction force arising from the many-body effects, presented in Fig.~\ref{fig:MBDvsPW_chains}, suggests that a substantial difference can be expected between MBD and PW models if the respective systems are allowed to deform or move dynamically. Indeed, first confirmations of such non-trivial force-deformation coupling, for the case of simplified short-range modeling, have been reported in previous studies \cite{Hauseux,PRL_colossal}. In the present work, we further study the effects of relaxation, but this time using the high-fidelity DFTB+MBD modeling framework to more accurately capture the short-range interplay between the DFTB and vdW contributions.

\begin{figure}[t]
\centering
    \subfloat[Distance between chains during debonding.]{\includegraphics[trim = 30mm 90mm 30mm 90mm, clip=true,width=0.49\textwidth]{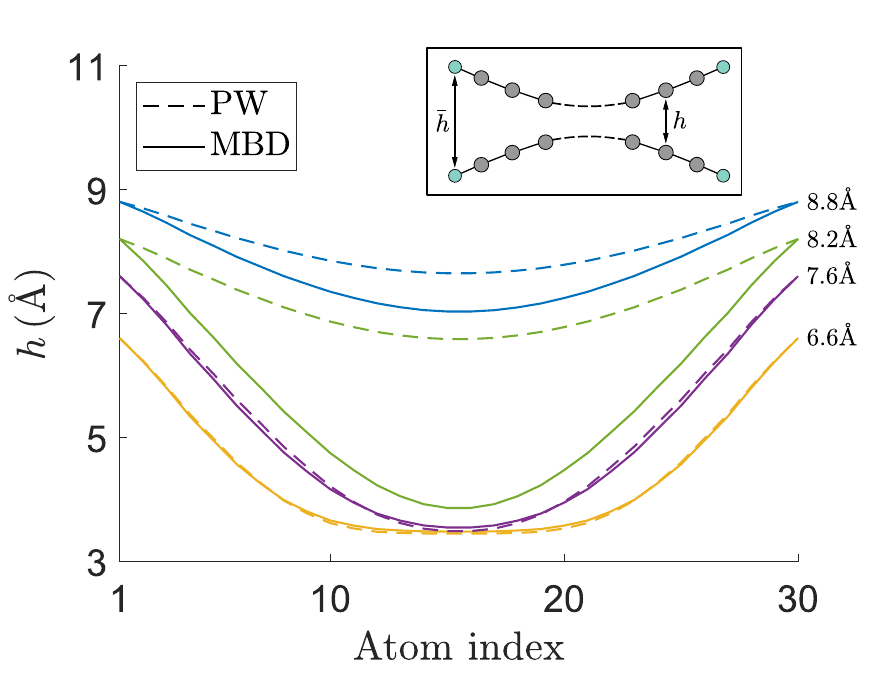}\label{fig:geo_bond}}\hfil
    \subfloat[Attraction force between chains.]{\includegraphics[trim = 30mm 90mm 30mm 90mm, clip=true,width=0.49\textwidth]{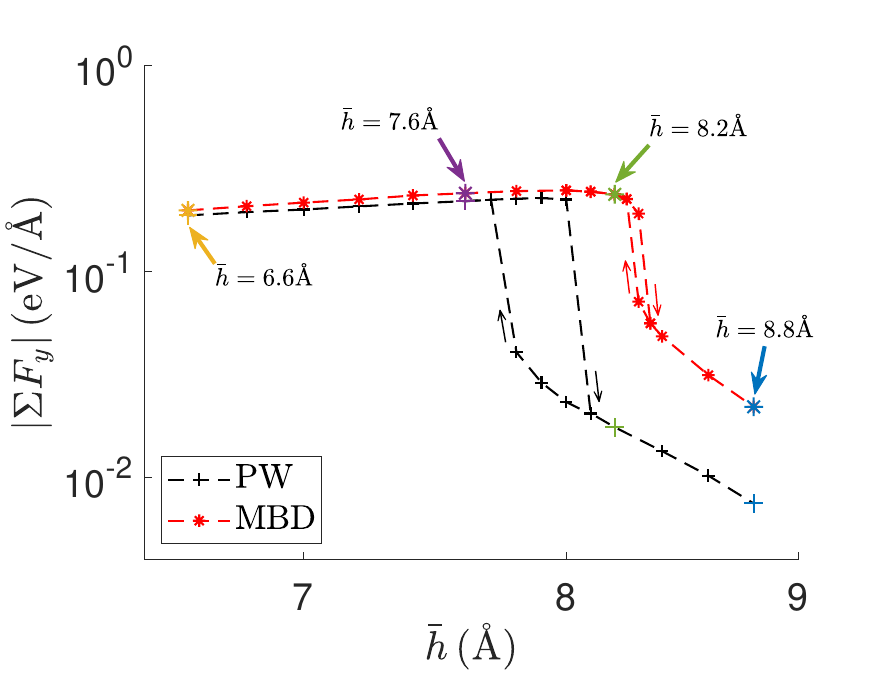}\label{fig:f_bond}}
    \caption{Bonding and debonding process between two deformable carbon chains. As depicted in the inset of (a), both chains are capped by hydrogen atoms (colored cyan) fixed at a distance of $\bar{h}$. In (a), the vertical gap profiles, $h$, between two chains are shown for selected instances of the debonding process. In (b), the force-separation hysteresis loops are shown for bonding-debonding cycles. The directions marked by black (PW) and red (MBD) arrows in (b) show the bonding/debonding directions. Colored arrows and markers in (b) correspond to respective cases in~(a).}
     \label{fig:CC_relaxation}
\end{figure}

In the flexible quasi-static case, we perform a single cycle of bonding/debonding between two equal chains composed of $N\text{C}_{1}=N\text{C}_{2}=28$ carbon atoms. In both chains, each of the edge carbon atoms is additionally capped by a hydrogen atom, which prevents the DFTB model predictions from being dominated by slowly decaying repulsive force coming from otherwise reactive endings. The chains are free to deform except for the hydrogen atoms that are fixed at a given inter-chain distance, $\bar{h}$. The process of bonding/debonding is subdivided into steps and is performed by gradually decreasing/increasing $\bar{h}$, respectively. At each step, the chain system deforms to its static force equilibrium, for which the inter-chain interaction force is computed. The results shown in Fig.~\ref{fig:CC_relaxation} reveal differences in responses between the two vdW models. Although both vdW models predict path-dependency of the process, the MBD hysteresis loop in Fig.~\ref{fig:f_bond} is relatively much smaller as compared to the PW model predictions. This can be explained by the fact that the net forces for MBD model are higher at longer distances, which leads to a more pronounced deformation and results in an earlier snap-in transition at bonding.

The rigid- and flexible quasi-static analyses of interacting chains revealed evident MBD effects as compared to the PW model predictions. Only at closer distances, the differences are shown to be much less pronounced, see for instance Figs.~\ref{fig:MBDvsPW_chains} and~\ref{fig:CC_relaxation} at the distances below $\bar{h}=7\,\text{\AA}$. This can be owed to the fact that at shorter distance ranges, the forces predicted by both vdW models are comparable and must additionally compete with exponentially increasing repulsive DFTB forces. 

Now, we will shift our focus to dynamics, which is expected to highlight more differences between vdW models. It has been shown that MBD yields significantly more accurate and complex dynamical properties compared to pairwise vdW models \cite{PRL_aspirin, Mario}, which can influence the prediction of macroscopic characteristics of material systems, such as thermal conductivity, rheological properties of fluids, and temperature-induced variations in elasticity. For the purpose of this paper, we will limit ourselves to a simple study that will illustrate basic dynamical effects induced by many-body interactions.

\begin{figure}[t]
\centering
\subfloat[Time-averaged $y$ deformation of the upper chain, $\Delta{}y$, at~$\bar{h}=8\,\text{\AA}$.]{\includegraphics[trim = 30mm 90mm 30mm 90mm, clip=true,width=0.48\textwidth]{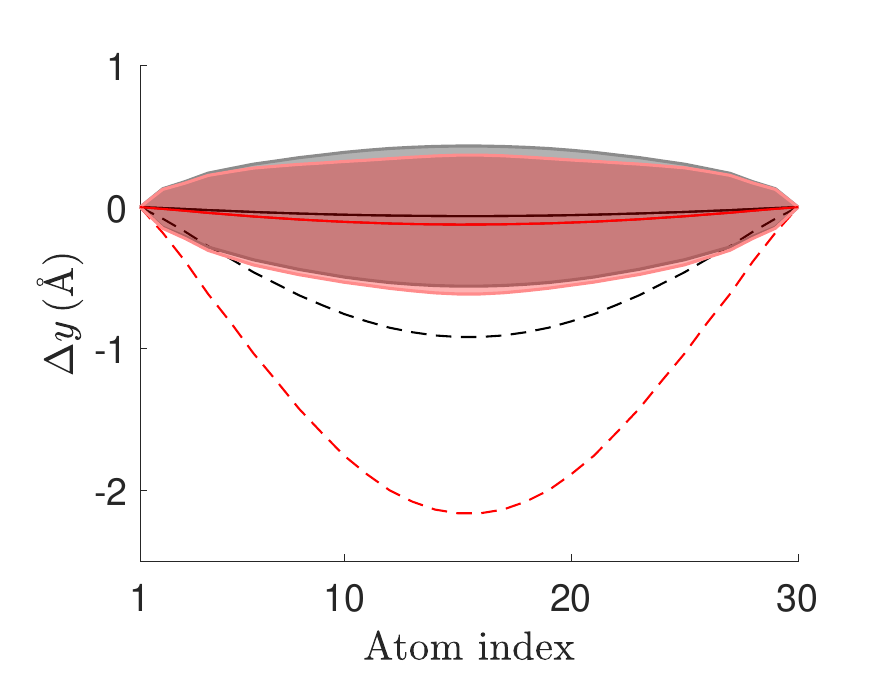}\label{fig:dynamic_position_y8a}}
\subfloat[Time-averaged $y$ deformation of the upper chain, $\Delta{}y$, at~$\bar{h}=7\,\text{\AA}$.]{\includegraphics[trim = 30mm 90mm 30mm 90mm, clip=true,width=0.48\textwidth]{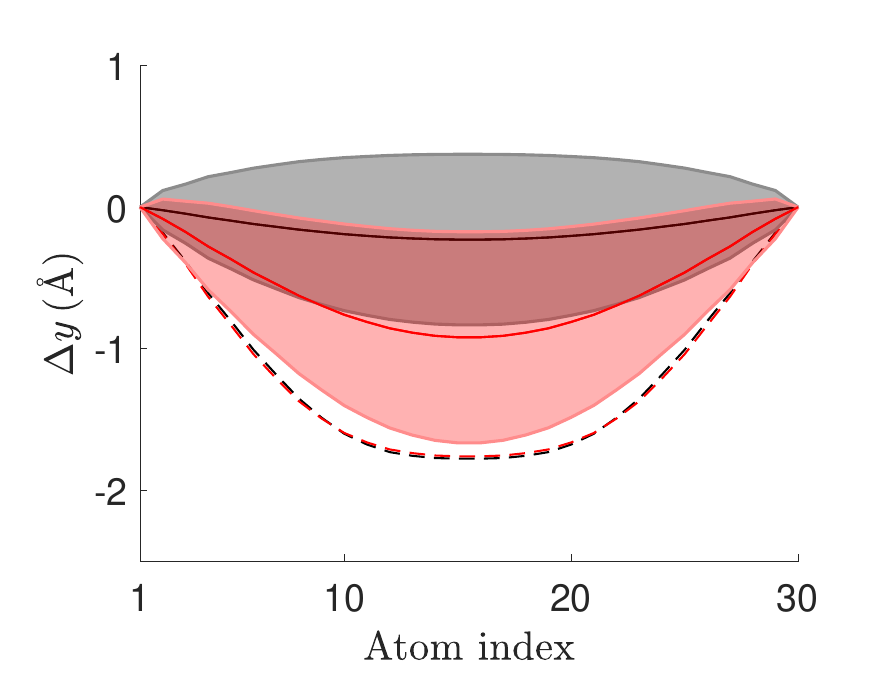}\label{fig:dynamic_position_y7a}}\hfill
\subfloat[Time-averaged $y$ deformation of the upper chain, $\Delta{}y$, at~$\bar{h}=6\,\text{\AA}$.]{\includegraphics[trim = 30mm 90mm 30mm 90mm, clip=true,width=0.48\textwidth]{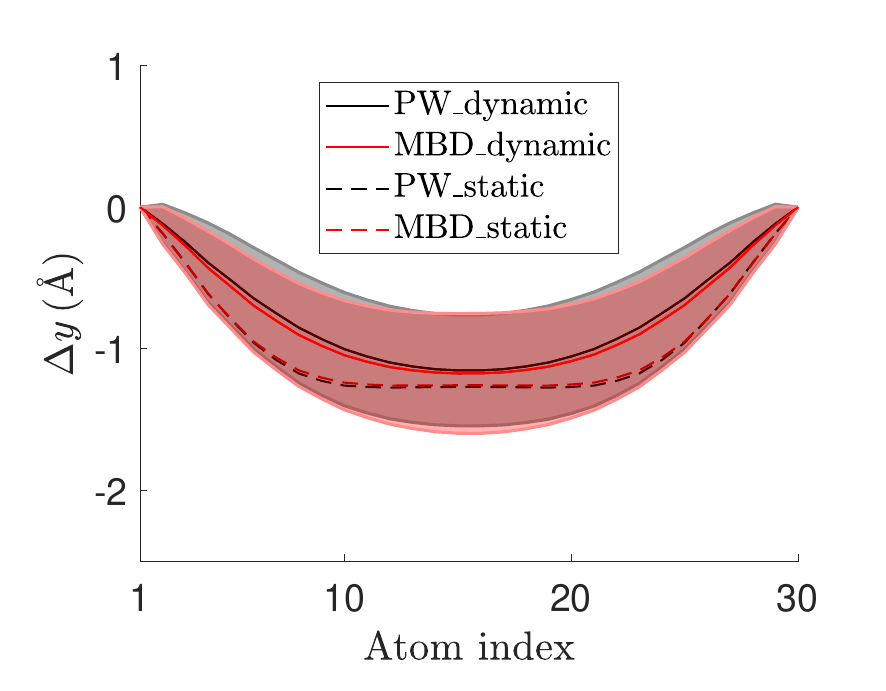}\label{fig:dynamic_position_y6a}}
\subfloat[Static forces vs time-averaged dynamic forces.]{\includegraphics[trim = 30mm 90mm 30mm 90mm, clip=true,width=0.48\textwidth]{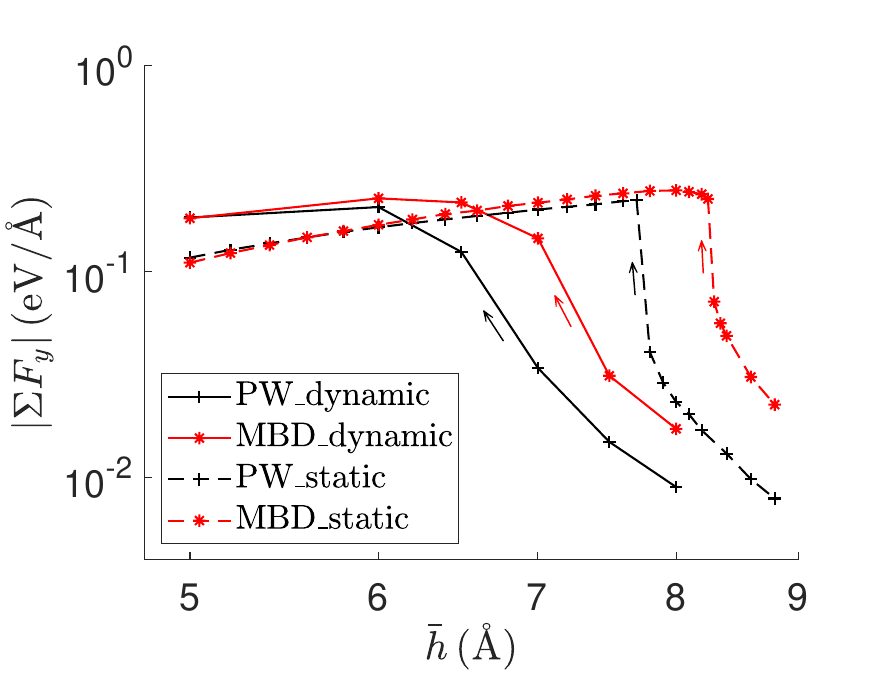}\label{fig:f_bond_dyn}}
    \caption{Dynamical bonding process between two deformable chains for the PW and MBD models. In (a)-(c), the time-averaged deformations in $y$ and respective standard deviations are shown for the upper chain at selected distances, with the corresponding static bonding results presented as references. In (d), The time-averaged force-bonding curve is shown, along with the static force-bonding curve for comparison. Red and black arrows in (d) mark the bonding direction.}
     \label{fig:corr_T300_L30_D10}
\end{figure}

In the dynamic analysis, we conduct molecular dynamics (MD) simulations of two interacting carbon chains at the temperature $300$\,K. Similarly to the quasi-static case, each chain consists of 28 carbon atoms and is capped at its boundaries by fixed hydrogen atoms. We analyze the system at different distances, $\bar{h}$, between the fixed hydrogen atoms, in the range $5$--$8\,\text{\AA}$, and compare the averaged results to the respective static deformation cases (see Fig.~\ref{fig:CC_relaxation} for a reference of the static cases). We use four different random seeds to parallelize the simulation. For each seed, after the initial run-up phase lasting $10^4$\,fs, the main phase of simulation continues until $10\times{}10^4$\,fs, during which the atoms' positions and forces are collected every 1\,fs time step. In~Fig.~\ref{fig:f_bond_dyn} we observe that the region of bonding-debonding shifted towards $\bar{h}= 7 \text{\AA}$ for the dynamic cases as compared to the static cases. At the same time, in the dynamic case, significant differences between responses predicted by MBD and PW models remained and were extended to a wider range of distances. These observations again emphasize the importance of using MBD as a high-fidelity model, which has been now demonstrated to be also valid in the dynamic setting.

\subsection{Single-wall carbon nanotube}
\label{sec:SWCN}

In this section, our framework is applied to study the mechanical responses of more complex atomic structures. As an example to study, we specifically choose the single-wall carbon nanotubes (SWCNTs or CNTs), which are known to possess exceptional properties due to their particular cylindrical arrangement of carbon atoms. The stiffness of around 1\,TPa \cite{CNT_E_experiment} and other potentially useful mechanical and electrical properties attracted attention from scientific communities and found a number of applications in industry \cite{CNT_composite_overview,CNT&Graphene_composite_science}.
Although various techniques have been developed so far for efficient modeling of CNTs \cite{Hossain_2018, REBO_CNT, Arash2014}, to our best knowledge, there is limited research that discusses aspects of efficiency versus accuracy in the context of DFTB and vdW models. 

We plan to utilize the predictive capabilities of the introduced DFTB-based framework to disentangle some of the effects that contribute to the mechanical responses of CNTs, and to demonstrate the pitfalls of simplified modeling. To do so, in Section~\ref{sec:harmonic} we introduce and calibrate harmonic models for SWCNTs, that will represent a class of simplified force-field methods. It will be then used in Section~\ref{sec:buckling} to showcase discrepancies of the simplified models with respect to the high-fidelity DFTB and vdW models.

\begin{figure}[h]
 \centering
\includegraphics[trim = 00mm 85mm 00mm 90mm, clip=true,width=0.6\textwidth]{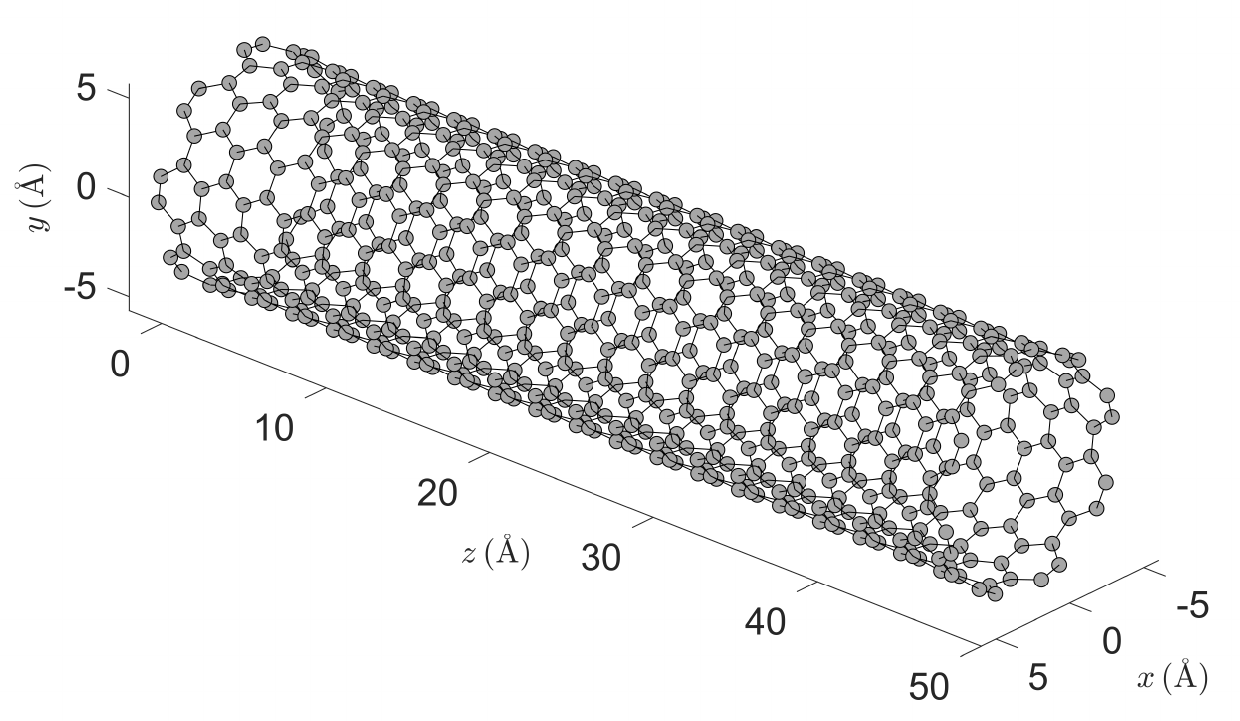}
    \caption{The SWCNT used in this section for the calibration of harmonic model and for the buckling tests. The number of carbon atoms $N_\text{C}=640$, radius $R_\text{CNT}=5.42 \text{\AA}$, and the reference length of the nanotube {$L=48\,\text{\AA}$}.}
    \label{fig:nanotube_geo}
\end{figure}

\subsubsection{Harmonic model and parameters calibration} \label{sec:harmonic}
Harmonic models can serve as a simplified empirical model for the DFTB interactions. They are widely utilized as classical force-field methods, such as AMBER \cite{AMBER}, GROMOS \cite{GROMOS05}, OPLS \cite{OPLS}, etc. In many applications, this popular and simple choice for modeling atomic interactions provides a sufficiently accurate approximation of covalent bonds in the context of molecular mechanics. More advanced models, such as REBO \cite{REBO} and SW \cite{sw}, incorporate nonlinear descriptions of covalent binding to enhance predictive power. However, they remain within the empirical domain and are inherently limited by the lack of short-range correlation and cumbersome parametrization.  

In the present work, we use a typical harmonic model, which can serve well for the purpose of illustration. Following the standard energy separation in MM, similarly to the energy separation introduced in Section~\ref{sec: E_SR E_LR}, the total energy $E^\text{tot}$ is split into covalent $E^\text{harmonic}$ and non-covalent $E^\text{vdW}$ parts. For three-dimensional nanostructures, such as CNTs, the harmonic part normally includes three independent terms, namely, bond stretching (2 atoms), bond bending (3 atoms), and bond torsion (4 atoms):
\begin{equation}
E^\text{harmonic} = \sum_i \dfrac{1}{2} k_r(r^i - r_0^i)^2 + \sum_j  \dfrac{1}{2} k_\theta(\theta^j - \theta_0^j)^2+ \sum_{l} \dfrac{1}{2} k_{\phi}(\phi^l - \phi_0^l)^2,
\label{eq:harmonic}    
\end{equation}
where $k_r$, $k_{\theta}$, and $k_{\phi}$ are bond force constants for stretching, bending, and torsion, respectively. $r$ is the bond length, $\theta$ is the bond angle, and $\phi$ is the torsion angle, also known as the dihedral angle. The subscript $0$ denotes reference values obtained from the initial relaxed configuration of the structure, which are allowed to be different for each bond (as denoted by the superscripts), assuring the energy minimum for the initial (relaxed) structure. Also, as there is no coupling between the three terms in Eq.~\eqref{eq:harmonic}, the bending and/or torsion terms can be optionally neglected, e.g., when analyzing simple atomic chain structures. 

The calibration of the three parameters in Eq.~\eqref{eq:harmonic} has been performed through fitting to the DFTB configurational energy of several thousand perturbed configurations, see the geometry in Fig.~\ref{fig:nanotube_geo}. 
The calibration resulted in the following fitted parameters: $k_r=35.0505\,\text{eV}/{\text{\AA}}^2$, $k_{\theta}=6.6069\,\text{eV}$, and $k_{\phi}=0.5361\,\text{eV}/\text{rad}^2$. 

\begin{remark} The presented calibration procedure was restricted to sampling the energy landscape around the equilibrium, which has its own drawbacks. The first one is that we are assuming the force to be linear on the energy evaluation variables, which is not the case for sufficiently large perturbations. Secondly, we are expecting our simplified model to be able to properly capture the energy landscape, which may not be the case even if we are well inside the linear regime. All of these combined can potentially influence the comparison between DFTB and the harmonic model at every range.
\end{remark}

\subsubsection{Buckling test}
\label{sec:buckling}
For the buckling test, we define the process as follows: First, we set up boundary conditions in which the two outermost layers of the nanotube (at its two ends) are fully fixed. At each compression step, $t$, the displacement increment $\Delta\bar{u}_{\text{z}}^t$ is applied to the right end of the nanotube. In the case of DFTB, we set $\Delta\bar{u}_{\text{z}}^t=0.027\,\text{\AA}$. For the harmonic model, the increment is $\Delta\bar{u}_{\text{z}}^t=0.1\,\text{\AA}$ and for cases without the dihedral contribution, we use $\Delta\bar{u}_{\text{z}}^t=0.05\,\text{\AA}$ to aid the minimization process, as the structure is less stable. Towards the end of the compression,  $\Delta\bar{u}_{\text{z}}^t$ can be further reduced to overcome convergence difficulties. 

At each compression step, $t$, we measure the the reaction force $F_{\text{z}}^t$, and compute the longitudinal strain of the nanotube $\varepsilon_{\text{zz}}^t=\bar{u}_{\text{z}}^t/L$. The instantaneous stiffness $E(\varepsilon_{\text{zz}}^t)$ is given by:
\begin{equation}
     E(\varepsilon_{\text{zz}}^t) = \frac{\Delta\sigma_{\text{zz}}^t}{\Delta\varepsilon_{\text{zz}}^t},
\end{equation}
where $\sigma_{\text{zz}}^t=F_{\text{z}}^t/S_{\text{face}}$ is the reaction stress on the constrained layer. The cross-section area of the nanotube $S_\text{face}$ is calculated by multiplying the nanotube circumference by the commonly assumed wall thickness $3.4\,\text{\AA}$, which equals the thickness of mono-layer graphene \cite{GUPTA20101049}.

\begin{figure}[h] 
 \centering
\subfloat[Harmonic models]{\includegraphics[trim = 30mm 90mm 30mm 90mm, clip=true,width=0.5\textwidth]{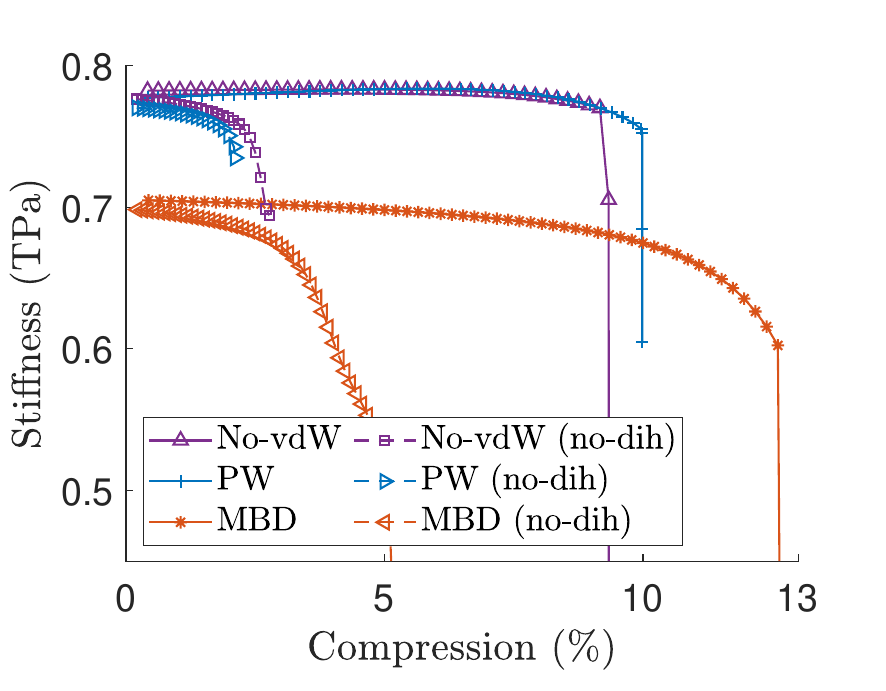}\label{fig:buckling_harmonic}}
\subfloat[DFTB model]{\includegraphics[trim = 30mm 90mm 30mm 90mm, clip=true,width=0.5\textwidth]{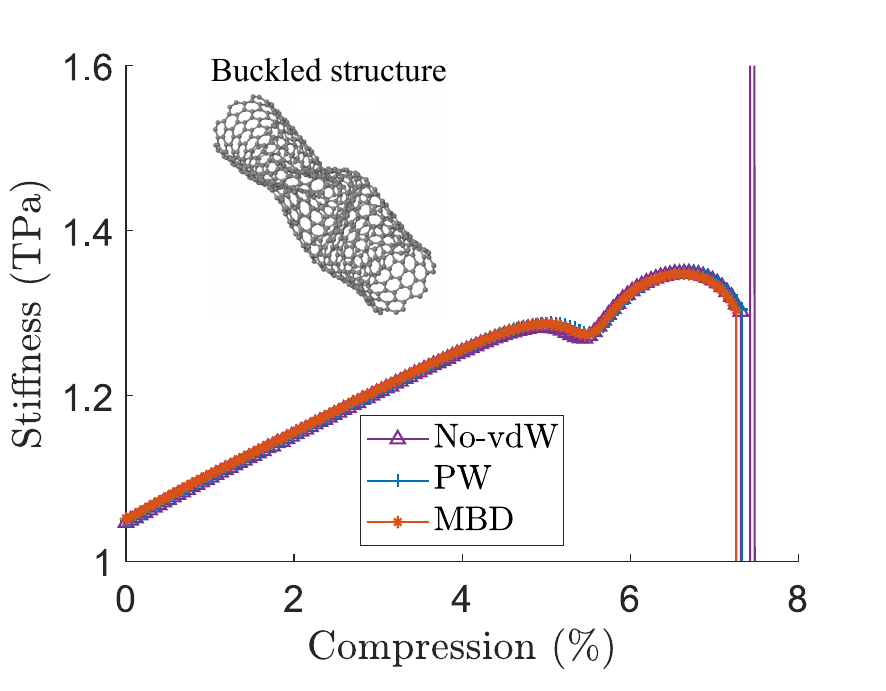}
\label{fig:Buckling_dftb_geom}}
    \caption{SWCNT Bbuckling results for the (a) harmonic models, and (b) DFTB model, for different vdW models. The global SWCNT buckling (see inset in (b)) only happens with DFTB, and is never observed for harmonic models.}\label{fig:buckling-plot}
\end{figure}

The results presented in Fig.~\ref{fig:buckling-plot} reveal completely different responses of harmonic-based models and DFTB-based models. Although in both cases, the values of initial stiffness fall not far from the expected 1\,TPa \cite{CNT_E_experiment}, it is clear that the simplified model fails to capture many aspects of the system under stress. First of all, the linear increase in stiffness observed in DFTB before buckling is not present in the harmonic model, as it assumes force linearity. Secondly, the initial stiffness value is underestimated, possibly due to its inability to properly capture the energy landscape near equilibrium. Furthermore, only local structural buckling is observed in the harmonic model, in contrast to the global buckling observed in DFTB, as shown in Fig.~\ref{fig:Buckling_dftb_geom}. This discrepancy again attributed to the lack of nonlinearity in the harmonic model. Finally, the harmonic model predicts meaningful differences with respect to vdW interactions in terms of buckling, which is not the case for DFTB. 

The fact that vdW effects do not significantly affect the responses of the DFTB model can be explained by the relatively high strength of covalent bonds which overrides much weaker vdW forces. The structure of SWCNTs do not leave any easily-activated DOF that could be controlled by vdW interactions. This important observation provided us with a clue on where to seek systems that rely more on vdW interactions, which leads us to polymer systems analyzed in Section~\ref{sec:UHMWPE}.

\begin{remark} When analyzing regions of rapid stiffness drops in the harmonic model (interpreted as buckling points), there is a significant qualitative difference depending on whether the model includes a dihedral term or not. As expected, the dihedral term adds resistance to buckling and brings the results of the harmonic model closer to DFTB. Additionally, it is worth mentioning that also MBD contributes to stabilizing the structure, which could be explained by its globally cohesive nature that regularizes the displacement field.
\end{remark}

\subsection{Ultra High Molecular Weight Polyethylene (UHMWPE)}
\label{sec:UHMWPE}

The two previous examples provided us with insights into how the mechanical responses of various simple systems can be driven by the interplay between short- and long-range interactions that are modeled by the DFTB+MBD framework. These observations naturally provoke the idea of analyzing three-dimensional systems that are less constrained than SWCNTs, thereby revealing higher sensitivity to vdW effects. In that respect, our choice is to study Ultra High Molecular Weight Polyethylenes (UHMWPEs) \cite{TAM20161,KURTZ20041}, which are three-dimensional arrangements of long polyethylene chains. As the interactions between these chains are primarily governed by vdW dispersion forces, they provide excellent systems for investigating qualitative differences induced by the choice of vdW models. 

\begin{figure}[h]
 \centering
\subfloat[Angle view]{\includegraphics[trim = 40mm 98mm 40mm 90mm, clip=true,width=0.49\textwidth]{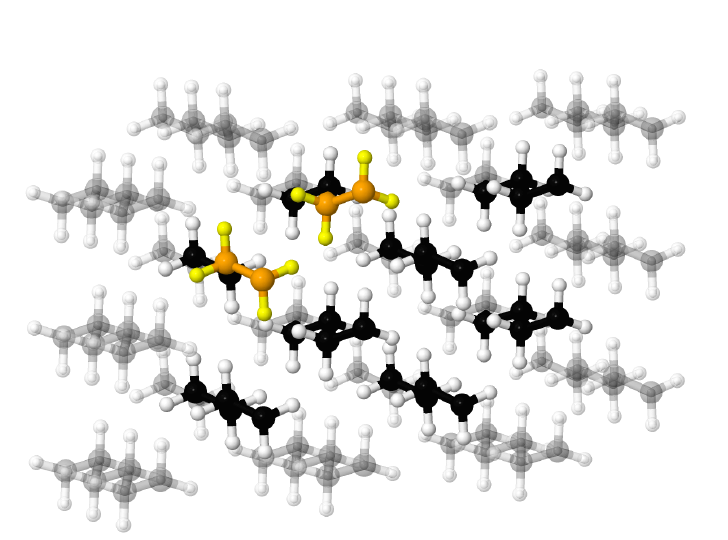}}
\subfloat[Front view]{\includegraphics[trim = 35mm 90mm 25mm 98mm, clip=true,width=0.49\textwidth]{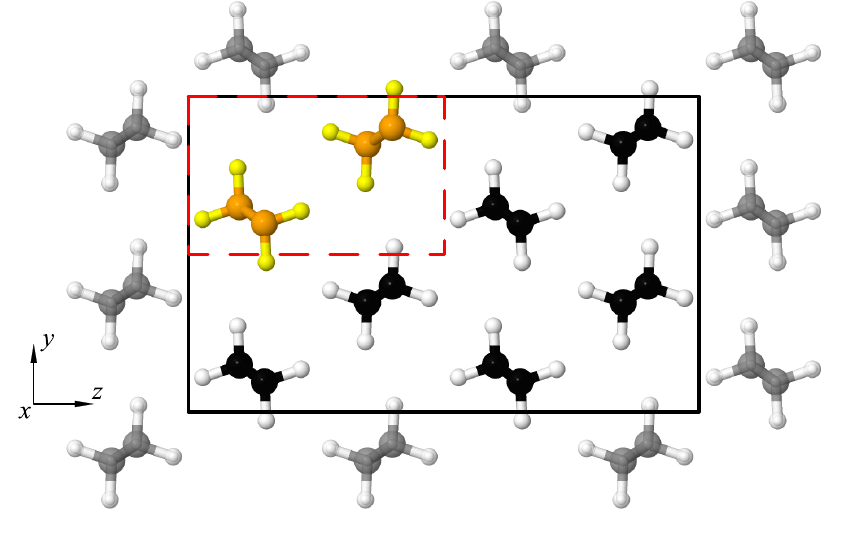}
\label{fig:UHMWPE_front}}
    \caption{Two different views of the UHMWPE crystal structure. Atoms inside a primitive unit cell (4 carbon and 8 hydrogen atoms) are colored yellow (H) and orange (C). The remaining atoms for a $2\times2\times2$ supercell are colored white (H) and black (C). Translucent atoms demonstrate the periodicity of the structure. In \ref{fig:UHMWPE_front}, the primitive unit cell is marked by a red dashed box, and the supercell is marked by a black solid box.} \label{fig:UHMWPE_structure}
\end{figure}

For the purpose of this study, we analyze polymer chains that are of \emph{crystalline} arrangements. Such crystal structure is typically represented by a unit cell that is repeating periodically over the space, see orange groups in Fig.~\ref{fig:UHMWPE_structure}. (In the 3-dimensional case, the crystal is defined by a $3\times3$ tensor $\boldsymbol{U}^{\text{cell}}$ that is composed of the three translation vectors for the unit cell.) The unit cells can form, so-called, \emph{supercells}, which are hexahedrons built of unit cells. A supercell consists of atoms having independent degrees of freedom (DOF), which are constrained by periodic boundary conditions. As such, larger supercells form less constrained periodic systems. 

Because of high computational costs, we limit ourselves to a quasi-static regime. We conduct two different loading scenarios. The first one is the compression in $x$ direction (longitudinal), affecting buckling behavior, see Section~\ref{sec: UHMWPE compression}. The second one is the elongation in $y$ direction (transversal), creating an irreversible reorganization of crystal structure, see Section~\ref{sec: UHMWPE elongation}. In both cases, we use the concept of periodic supercells, and the boundary conditions are applied to the boundaries of periodic supercells, not the individual atoms. The sizes of periodic supercells are chosen large enough to demonstrate effects of interest.

\begin{remark}
    As a preliminary step in each case analyzed, the periodic supercells are first fully relaxed. This is accomplished by applying zero-stress Neumann boundary conditions to the periodic boundaries of the supercell and then allowing the system to converge. Such relaxed structures are characterized by expected zero reaction forces at the periodic boundaries at the beginning of the respective compression/tension tests.
\end{remark}

\subsubsection{Compression test}
\label{sec: UHMWPE compression}

\begin{figure}[th]
 \centering
\subfloat[Maximum principal stress]{\includegraphics[trim = 30mm 90mm 30mm 90mm, clip=true,width=0.49\textwidth]{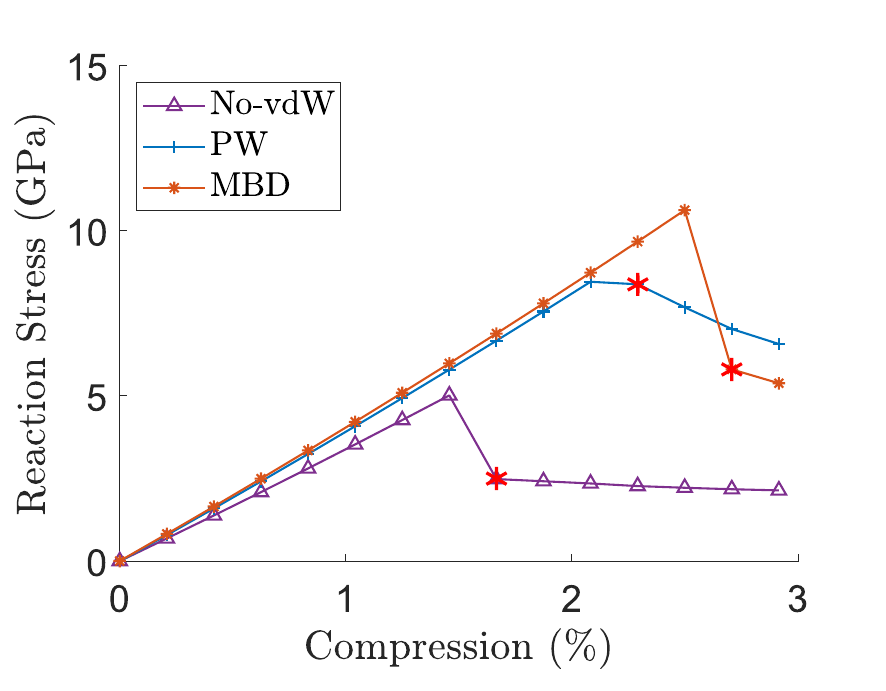}}
\subfloat[$L^2$ norm of the 2 transversal principal stresses]{\includegraphics[trim = 30mm 90mm 30mm 90mm, clip=true,width=0.49\textwidth]{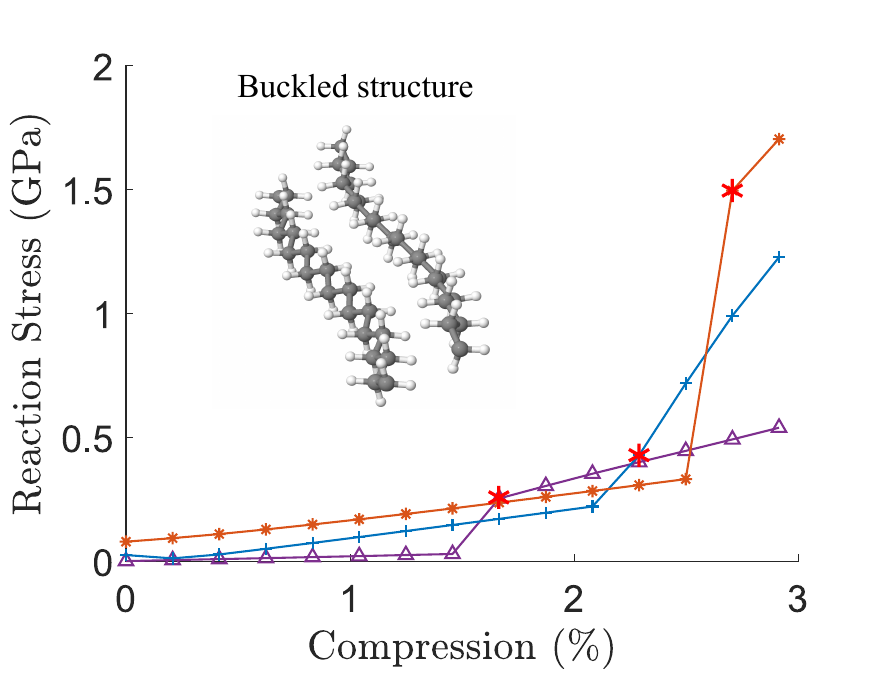}\label{fig:UHMWPE_buckling_lateral}
}
    \caption{UHMWPE compression test for different vdW models. In (a), the maximum principal stress represents the reaction in the direction of compression. In (b), the transversal reaction stress is presented as the $L^2$ norm of two minor principal stresses. The red asterisks in both figures indicate the initial buckling point of each case. In the inset in (b), the buckled structure at $3\%$ compression is presented (it is very similar among all three cases).}
    \label{fig:UHMWPE_buckling}
\end{figure}

As aforementioned, the compression test is performed in the longitudinal direction. We choose the $N_{\text{x}}\times{}N_{\text{y}}\times{}N_{\text{z}}=10\times{}1\times{}1$ supercell that involves two long polyethylene chains. Such increased aspect ratio of the supercell promotes earlier buckling behavior of the structure (see the inset in Fig.~\ref{fig:UHMWPE_buckling_lateral}).

The buckling of the periodic structure is achieved by compressing the supercell in the $x$ direction while having the remaining DOFs of the supercell fixed. Technically, we decrease the $U^{\text{cell}}_{xx}$ component of the $U^{\text{cell}}$ tensor and keep its remaining components fixed. The compression is performed until $3\%$ of reduction, with the step size, $\Delta U^{\text{cell}}_{xx}$, of $-0.053\,\text{\AA}$ (i.e., $0.2\%$ of compression). At each step of compression, we retrieve the stress tensor that corresponds to the reaction at the constrained supercell, which can be interpreted as the macroscopic stress. We analyze the principal stresses, with the maximum principal stress corresponding to the loading (longitudinal) direction and the two remaining minor principal stresses corresponding to the reactions in the transversal direction. 

The results are presented in Fig.~\ref{fig:UHMWPE_buckling}. The first observation is that the predictions strongly depend on vdW model, which demonstrates a qualitative difference of the UHMWPE systems as compared to SWCNTs (see Fig.~\ref{fig:Buckling_dftb_geom}).
The inclusion of vdW interactions delays buckling and predicts a slightly stiffer UHMWPE, despite the fact that there is no significant difference observed in the buckled structure among the three cases. For the transversal reaction, vdW interactions exhibit a pronounced impact on the post-buckled phases, leading to a rapid increase in reaction stresses. Notably, the MBD model predicts higher transversal stresses of the buckled structure, which highlights the significance of many-body effects on the non-covalent bonding. These observations are consistent with our hypothesis that in multi-molecule systems with lower stiffness, the vdW dispersion interaction and its many-body nature play crucial roles, and therefore need to be treated carefully by using high-fidelity models. 

\subsubsection{Elongation test}
\label{sec: UHMWPE elongation}

In the elongation test we consider two different sizes of supercells, see Fig.~\ref{fig:UHMWPE_structure}. The smallest $1\times1\times1$ supercell is composed of a single primitive unit cell, which constrains the system to 12 atoms with 3 DOFs each. We also use the $2\times2\times2$ supercell, which increases the number of independent atoms by the factor of 8, allowing us to study more complex reorganization of polymer chains. 

The elongation is performed by gradually increasing the $U^{\text{cell}}_{yy}$ component of the $U^{\text{cell}}$ tensor while allowing other components to relax. The increments $\Delta U^{\text{cell}}_{yy}$ are $0.027\,\text{\AA}$ and $0.053\,\text{\AA}$ for the small and large supercells, respectively, which assures relatively the same increments in both cases. At each increment, we retrieve the $yy$ component of the reaction stress tensor. (Note, that due to the relaxation of the remaining components of the deformation tensor, the reaction in the $y$ direction is the only non-zero component of the stress tensor.) 

\begin{figure}[th]
 \centering
\subfloat[$1\times 1 \times 1$ supercell]{\includegraphics[trim = 30mm 90mm 30mm 90mm, clip=true,width=0.49\textwidth]{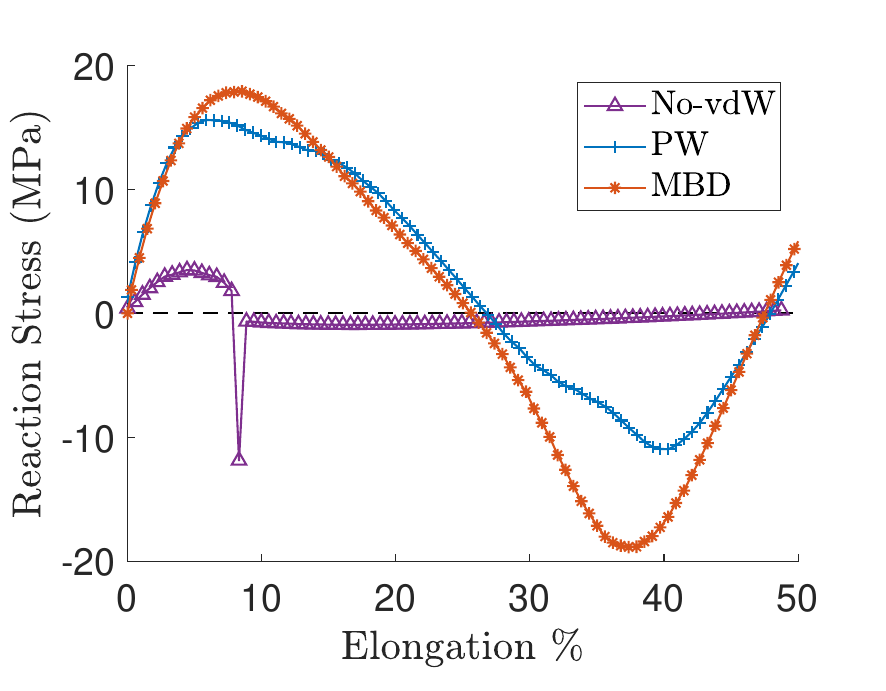}}
\subfloat[$2\times 2 \times 2$ supercell]{\includegraphics[trim = 30mm 90mm 30mm 90mm, clip=true,width=0.49\textwidth]{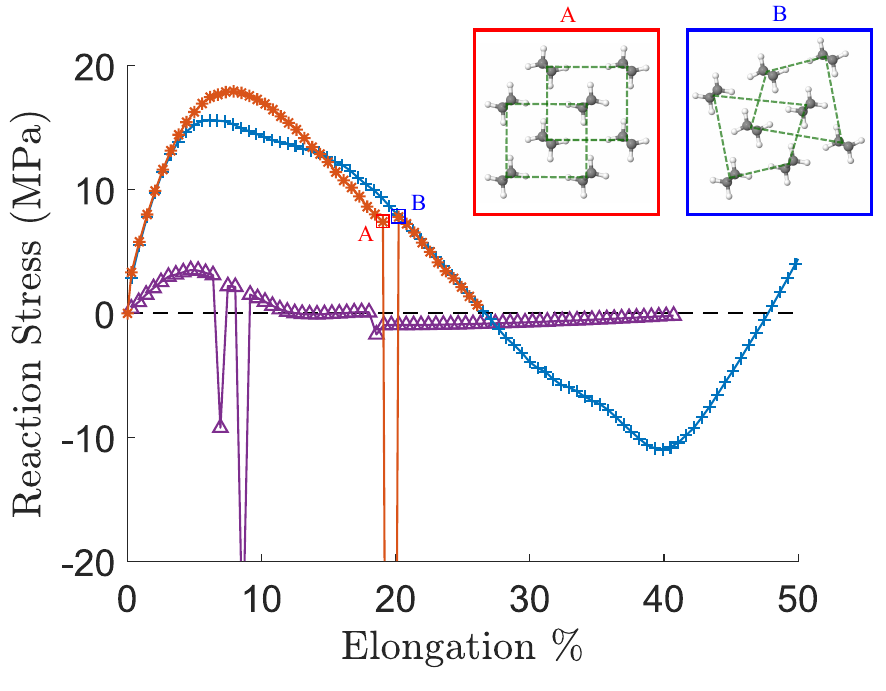}\label{fig:UHMWPE_stretch_super}
}
    \caption{Evolution of reaction stress during the elongation of UHMPWE structures in the $y$ direction for two different supercells and three different vdW cases. The A and B insets in (b) show the abrupt transition of UHMPWE microstructure between respective points A and B on the load curve in MBD plot. The green dashed lines in the insets connect the polyethylene chains of the same orientation, which makes it easier to observe the structural change. }
    \label{fig:UHMWPE_stretch}
\end{figure}

The results are presented in Fig.~\ref{fig:UHMWPE_stretch}. The first clear observation is that vdW interactions are indeed crucial to be included in the modeling of UHMWPE. The No-vdW model predicts relatively early breakage of the structure, while the inclusion of vdW interactions maintains structural integrity and introduces non-trivial effects throughout the entire elongation range. The second observation concerns stress jumps visible in the No-vdW and MBD plots. These jumps indicate phase transitions involving significant changes in both the relative positions and orientations of the polyethylene chains. In the case of the No-vdW model, the jumps indicate moments of structural breakage, whereas in the case of the MBD model, the jump indicates the moment of phase transition. The third observation relates to differences between the predictions of the PW and MBD models. The most significant qualitative difference is the phase transition predicted only by the MBD model, as shown in Fig.~\ref{fig:UHMWPE_stretch_super}. Even more interesting is that the stresses at point B in Fig.~\ref{fig:UHMWPE_stretch_super} are almost identical for both vdW models, whereas the microstructure predicted by the MBD model undergoes the phase change, and the one predicted by the PW model remains unchanged. It is also worth noting that the phase change only occurs for the larger supercell, which again highlights the important role of many-body effects in systems with more ``easily-activated'' degrees of freedom, such as those with larger supercells or longer chains.

Even though we limited ourselves to a quasi-static regime, the observations made for dynamic chain-to-chain interactions in Section~\ref{sec:vdW_chain} suggest that the effects observed for UHMWPE can remain or even be amplified in the fully dynamic regime. In fact, in the case of polymer melts--where chains are not tethered at the ends as in our simulations--the dynamic freedom of the chains introduces a multitude of configurations that may frequently fall within the regime where predictions of MBD and PW models diverge significantly. The cooperative motions that are fundamentally captured by the MBD model, and are absent in the PW model, could lead to alternative entanglement dynamics and mesoscopic structures in polymer melts. These structures, in turn, are pivotal in determining the material's macroscopic properties such as viscosity, mechanical strength, and thermal conductivity.

\section{Conclusions and future work}
\label{sec:conclusion}

In this paper, we studied the importance of including quantum-based models to accurately predict the mechanical properties of materials. For this purpose, we used a high-fidelity modeling framework, DFTB+MBD, that combines the Density Functional Tight Binding (DFTB) method with van der Waals (vdW) dispersion interactions, including both many-body dispersion (MBD) and pairwise methods. This framework is designed to address the limitations of classical force fields and simplified models in capturing the complex interactions that occur at the atomic and molecular scales, particularly in systems where quantum effects are significant.

Through a series of benchmark studies on various molecular systems, we demonstrated that the DFTB+MBD framework can be a suitable high-fidelity validation toolkit, able to reveal spurious or misleading effects predicted by selected simplified models. In particular, we showed appreciable differences in stiffness, reaction forces, and critical points of structural change. These differences were caused in part by the underlying linearity of the simplified harmonic models, which is inaccurate when considering sufficiently large perturbations, but also by the effects of many-body dispersion interactions that are not accurately predicted by the simplified pairwise models. 

Our findings shed light on how the responses of different material systems are influenced by quantum-scale effects. We underscored the essential role of the many-body nature of vdW interactions, as captured by the MBD method, particularly in systems with extended non-bonded interactions. A notable example is Ultra High Molecular Weight Polyethylene (UHMWPE), which exhibits sensitivity to vdW effects. While our research is currently limited to the quasi-static regime and relatively small supercell sizes, we identify UHMWPE as a promising candidate for future studies. For example, dynamic simulations involving larger supercells could enhance the predictive capabilities for transition points, viscosity, elasticity, and other macroscopic properties of polymer melts.

Lastly, we presented an open-source toolkit \cite{QuaCrepo1} to facilitate the implementation of the DFTB+MBD framework, thereby making numerical implementations and molecular simulations more accessible to the engineering community. This framework and toolkit offer a practical approach for investigating phenomena at the microscale with high fidelity. We anticipate that they will support future research in effective ab-initio large-scale and multiscale modeling. Specifically, we aim to focus our future efforts on developing a more general workflow for constructing effective simplified models, which will extend the applicability of the framework to engineering-level systems.

\section*{Acknowledgments}
We are grateful for the support of the Luxembourg National Research Fund (C20/MS/14782078/QuaC). We extend our thanks to Mario Galante for the many fruitful discussions we had about the topics of this paper. The calculations presented in this paper were carried out using the HPC facilities of the University of Luxembourg.


 \bibliographystyle{elsarticle-num}
 \bibliography{cas-refs}






\end{document}